\documentclass[preprint, 11pt, authoryear]{elsarticle}
\pdfoutput=1
\usepackage{amsmath,amsthm}%
\usepackage{amsfonts, color}%
\usepackage{amssymb}%
\usepackage{natbib}
\usepackage{graphicx}
\usepackage{verbatim} 
\usepackage[T1]{fontenc} 
\usepackage[utf8]{inputenc} 
\usepackage{booktabs}
\usepackage{comment}
\usepackage{multirow}
\usepackage[dvipsnames]{xcolor}
\usepackage{xr-hyper}
\RequirePackage[colorlinks,citecolor=blue,urlcolor=blue]{hyperref}
\usepackage[margin=1in]{geometry}

\usepackage{fleqn}
\usepackage{endfloat}

\newtheorem{theorem}{Theorem}

\newtheorem{definition}{Definition}

\DeclareMathOperator{\sgn}{sgn}

\newcommand{\bdsm}{\boldsymbol}

\begin{document}

\begin{frontmatter}

	\title{Interval-Valued Kriging Models for Geostatistical Mapping with Imprecise Inputs}

 	\author[1]{Brennan Bean\corref{cor1}}
	\ead{brennan.bean@usu.edu}

	\author[1]{Yan Sun}
	\ead{yan.sun@usu.edu}

	\author[2]{Marc Maguire}
	\ead{marc.maguire@unl.edu}

	\cortext[cor1]{Corresponding author}

    \address[1]{Department of Mathematics and Statistics, Utah State University, 3900 Old Main Hill, Logan, UT 84322}
	\address[2]{Durham School of Architectural Engineering and Construction, University of Nebraska - Lincoln, 1110 S. 67th Street, Lincoln, NE 68588}

\begin{abstract}
Many geosciences data are imprecise due to various limitations and uncertainties in the measuring process. One way to preserve this imprecision in a geostatistical mapping framework is to characterize the measurements as intervals rather than single values. To effectively analyze the interval-valued data, this paper proposes and develops interval-valued kriging models based on the theory of random sets and a generalized $L_2$ metric. These models overcome the mathematical difficulties of a previous development and are computationally more feasible. Numerical implementation of our interval-valued kriging is provided using a penalty-based constrained optimization algorithm. An application to the prediction of design ground snow loads in Utah, USA, is presented that demonstrates the advantages of the proposed models in preserving crucial sources of uncertainty towards a more efficient risk-based designing scheme. 
\end{abstract}


\end{frontmatter}

Kriging models have become ubiquitous in the last 70 years, with entire books devoted to explaining their many applications \citep{Goovaerts1997}. These models continue to have modern relevance in a wide array of disciplines \citep{Shtiliyanova2017, Mao2018, Jin2018}. The foundation of the kriging paradigm rests upon the theory of regionalized variables laid out by \cite{Matheron71}, which are random functions that vary subject to an underlying spatial process within a geographical region. Matheron shows how proper considerations of spatial structure eliminate systematic errors and increase the precision of predictions as compared to methods which assume no spatial structure. This spatial structure is often characterized by defining the covariance between observations as a function of their deviation in location. In a less restrictive setting, kriging models need only assume that the variance of the \textit{difference} between two observations, called the semivariance, can be characterized as a function of location deviation. In both settings, the kriging predictions rely on a well-defined, real-valued function that measures the similarity between observations as a function of location deviation. 

These kriging models with their associated covariance/semivariance functions necessarily assume that observations in the field are known, precise values. This assumption limits the direct applicability of these models in situations where the observations are imprecise due to limitations in measurement technology. For example, air quality measurements can be limited in their abillity to detect trace levels of a pollutant \citep{Trivikrama1991}. Thus, a ``zero-valued'' observation may actually be an imprecise value falling somewhere between zero and the minimum detection limit. Imprecision is also prevalent in remote sensing data, where reconciling the imprecision that exists due to differences in the spectral response of different cameras is still an open problem \citep{Mckee2018}. Indeed, the recent explosion in the diversity and amount of remote sensing data brings with it the need for methods that can appropriately consider a spectrum of data precision from multiple sources. Traditional kriging methods require this imprecision to be either ignored or handled indirectly. Some variants of kriging characterize imprecision of the data through a series of indicator functions \citep{Hohn1998}, or seek to indirectly model the parameters of a distribution intended to characterize local uncertainty \citep{Goovaerts1997}. In either case, these approaches fall short of directly including the measurement imprecision in model formulations and predictions.

Several attempts to directly incorporate data imprecision as intervals within the kriging framework occurred in the late 1980s (see \cite{Loquin2010} for a relatively comprehensive review). \cite{Diamond88} proposed an interval-valued kriging where interval inputs where used to make interval predictions. This model is a direct extension of the point-valued kriging \citep{Matheron63, Matheron71} with an interval-valued random function and a proposed covariance structure for intervals, as explored in further detail in Section \ref{intkrige:models}. Later, \cite{Diamond89} extended his initial attempt to a fuzzy kriging, which allowed membership degrees for each interval \citep{Zadeh1965}. At about the same time, \cite{Bardossy90a, Bardossy90b} proposed a variant of fuzzy kriging where uncertainties in the variogram parameters were characterized as fuzzy intervals. Little attention was given to these models in the subsequent 30 years until \cite{Bandemer2000} attempted a Bayesian extension of Diamond's fuzzy kriging, and \cite{Loquin2012} attempted an algorithmic extension of Bardossy's fuzzy kriging. This 30 year lack of attention to these models is likely attributed to the computational difficulties of each framework given the computational limitations of the time. Another limitation of these models was the absence of a well-defined notion of covariance/semivariance between intervals. This severely restricts the applicability of these models as the covariance/semivariance cannot be directly estimated from the data. 

Thus, there remains the need for data-driven, computationally feasible, kriging models that can directly accommodate data of varying precision. For this reason, this paper proposes a modification to Diamond's interval-valued kriging based on the recent developments of the random set theory and a generalized $L_2$ distance. These new models overcome many of the mathematical and computational difficulties of previous interval-valued kriging attempts through the use of a well-established, real-valued covariance between intervals. The resulting predictions offer two distinct measures of uncertainty, as uncertainty due to spatial configuration is modeled by the kriging variance, while uncertainty due to imprecise input is modeled by the predicted interval radius. The numerical implementations of these models in R 3.6.1 \citep{R2019} leverage existing geospatial workflows in the \texttt{sp} and \texttt{gstat} packages \citep{gstat2016, gstat2004, sp1, Bivand2013} and are available on CRAN \citep{Bean2019-ik}. The feasibility and utility of these models are demonstrated through interval-valued kriging predictions of design ground snow loads in Utah, USA. 

The rest of this paper proceeds as follows. Section \ref{prelim} reviews the random sets framework underlying the proposed interval-valued kriging models introduced in Section \ref{intkrige:models}. Section \ref{numerics} describes the details of the numerical implementation while Section \ref{body3:application} presents the application of our interval-valued kriging to the design ground snow load prediction problem. We give concluding remarks in Section \ref{reflections} and provide technical proofs of the theorems in the Appendix.

\section{Random sets preliminaries}\label{prelim}
Denote by $\mathcal{K}\left(\mathbb{R}^d\right)$ or $\mathcal{K}$ the collection of all non-empty compact subsets of $\mathbb{R}^d$. The Hausdorff metric $\rho_H$
\[
  \rho_H\left(A,B\right)=\max\left(\sup\limits_{a\in A}\rho\left(a,B\right), \sup\limits_{b\in B}\rho\left(b,A\right)\right),\ \forall A,B\in\mathcal{K},\]
where $\rho$ denotes the Euclidean metric, 
defines a natural metric in $\mathcal{K}$. As a metric space, $\left(\mathcal{K}, \rho_H\right)$ is complete and separable \citep{Debreu67}.  In the space $\mathcal{K}$, a linear structure can be defined by Minkowski addition and scalar multiplication as
\[\label{def:int-linear}
  A+B=\left\{a+b: a\in A, b\in B\right\},\ \ \ \ \lambda A=\left\{\lambda a: a\in A\right\},\ \ 
  \forall A, B\in\mathcal{K},\ \  \lambda\in\mathbb{R}.
\]
Note however that $\mathcal{K}$ is not a linear space (or vector space) as there is no inverse element of addition. Let $(\Omega,\mathcal{L},P)$ be a probability space. A random compact set is a Borel measurable function $A: \Omega\rightarrow\mathcal{K}$, $\mathcal{K}$ being equipped with the Borel $\sigma$-algebra induced by the Hausdorff metric. If $A(\omega)$ is convex almost surely, then $A$ is called a random compact convex set \citep{Molchanov05}. The collection of all compact convex subsets of $\mathbb{R}^d$ is denoted by $\mathcal{K}_{\mathcal{C}}\left(\mathbb{R}^d\right)$ or $\mathcal{K}_{\mathcal{C}}$. 

Particularly, $\mathcal{K}_{\mathcal{C}}(\mathbb{R})$ contains all the non-empty bounded closed intervals in $\mathbb{R}$ and a measurable function that maps $\Omega$ to $\mathcal{K}_\mathcal{C}\left(\mathbb{R}\right)$ is called a random interval. From now on, an element in $\mathcal{K}_{\mathcal{C}}(\mathbb{R})$ will be denoted by $[x]$, whose lower/upper bounds and center/radius are denoted by $x^L$/$x^U$ and $x^C$/$x^R$, respectively. Bold letters denote vectors and random versions are denoted by capital letters. For example, $[\bdsm{x}]=\left[[x_1], \cdots, [x_p]\right]^T$ denotes a $p$-dimensional hyper interval and its random version is denoted by $[\bdsm{X}]$. The expectation of a random compact convex random set $A$ is defined by the Aumann integral of set-valued function \citep{Artstein75, Aumann65} as $E\left(A\right)=\left\{E\xi:\xi\in A \text{ almost surely}\right\}$, which for a random interval $[X]$ is $E\left([X]\right)=[E\left(X^L\right),E\left(X^U\right)]$. 

For interval-valued data analysis, the measure of distance is a critical issue. According to the embedding theorems \citep{Radstrom52, Hormander54}, $\mathcal{K}_{\mathcal{C}}$ can be embedded isometrically into the Banach space $C(S)$ of continuous functions on the unit sphere $S^{d-1}$, which are realized by the support function of $X\in\mathcal{K}_{\mathcal{C}}$. 
Therefore, a compact convex set can be represented by its support function $s_{X}$ and $\rho_2\left(X, Y\right):=\|s_X-s_Y\|_2$, $\forall X, Y\in\mathcal{K}_\mathcal{C}$, defines an $L_2$ metric on $\mathcal{K}_\mathcal{C}$. It is known that $\rho_H$ and $\rho_2$ are equivalent metrics, but $\rho_2$ is more preferred for statistical inference, due to many of its established properties \citep{Korner95, Korner97}. The $\rho_2$-metric for an interval $[x]$ has the particularly simple form
$\|[x]\|^2_2= (1/2)\left(x^L\right)^2+(1/2)\left(x^U\right)^2=\left(x^C\right)^2+\left(x^R\right)^2$
and the $\rho_2$-distance between two intervals is 
\[\label{def:p2}
\rho^2_2\left([x], [y]\right)= (1/2)\left({x}^L-{y}^L\right)^2+(1/2)\left({x}^U-{y}^U\right)^2= \left(x^C-y^C\right)^2+\left(x^R-y^R\right)^2. 
\]
A more general metric for $\mathcal{K}_{\mathcal{C}}(\mathbb{R})$ was proposed \cite{Gil01} which essentially takes form 

\begin{equation}\label{def:w2}
   \rho_W^2\left([x],[y]\right)
   =\left(x^C-y^C\right)^2+\left(x^R-y^R\right)^2\int_{[0,1]}\left(2\lambda-1\right)^2dW(\lambda),
\end{equation} 
where $W$ is any non-degenerate symmetric measure on $[0, 1]$. This allows for weighting between the center and radius. Separately, for a more general space, \cite{Korner01} proposed another $L_2$ metric, which when restricted to $\mathcal{K}_{\mathcal{C}}(\mathbb{R})$ is
\[
\rho^2_K([x], [y])=\sum_{(u, v)\in S^{0}\times S^{0}}\left(s_{[x]}(u)-s_{[y]}(u)\right)\left(s_{[x]}(v)-s_{[y]}(v)\right)K(u, v),
\]
where $K$ is a symmetric positive definite kernel. It can be represented by the upper/lower bounds as
\begin{eqnarray*}\label{def:k2-ul}
  \rho^2_K([x], [y]) &=& \left(x^U-y^U\right)^2K(1,1)+\left(x^L-y^L\right)^2K(-1,-1) \nonumber \\
  &&- \left(x^U-y^U\right)\left(x^L-y^L\right)\left[K(1,-1)+K(-1,1)\right], 
\end{eqnarray*}
or equivalently in the center-radius form as 
\begin{equation}\label{def:k2-cr}
  \rho^2_K([x], [y])=A_{11}(x^C-y^C)^2+A_{22}(x^R-y^R)^2+2A_{12}(x^C-y^C)(x^R-y^R),
\end{equation}
where 
\begin{eqnarray*}
  &&A_{11}=K(1,1)+K(-1,-1)-\left[K(1,-1)+K(-1,1)\right],\\
  &&A_{22}=K(1,1)+K(-1,-1)+\left[K(1,-1)+K(-1,1)\right],\\
  &&A_{12}=A_{21}=K(1,1)-K(-1,-1).
\end{eqnarray*}
Apparently, when $K$ is symmetric positive definite, so is $A$. Thus the essence of $\rho_K$ lies in its further generalization of $\rho_W$ that takes into account the interaction between the center and the radius. 

\section{The interval-valued kriging}\label{intkrige:models}

Let $Z(\bdsm{x})$ be a random function over a geographical region $\mathcal{R}$, typically realized by $\mathbb{R}^2$ or $\mathbb{R}^3$, where $\bdsm{x}$ represents a point (i.e., coordinates) in $\mathcal{R}$. An important kriging assumption is the second-order stationarity of $Z(\cdot)$, which means either covariance or variogram stationarity.  The covariance stationarity, for example, is defined by:
\begin{enumerate}
\item $E(Z(\bdsm{x})) = m$ is constant and independent of $\bdsm{x}\in\mathcal{R}$,
\item $Cov(Z(\bdsm{x}),Z(\bdsm{x} + \bdsm{h}) = C(\bdsm{h})$ exists and is independent of $\bdsm{x}$. 
\end{enumerate}
Under this assumption, the kriging estimator is defined as 
\[\hat{Z}(\bdsm{x}^*) = \sum_{\alpha = 1}^n\lambda_\alpha Z(\bdsm{x}_\alpha),\]
where $\bdsm{x}_\alpha$ represents the location of the surrounding measurements and the $\lambda_\alpha$ are selected to minimize
\[\text{Var}(\hat{Z}(\bdsm{x}^*) - \hat{Z}(\bdsm{x}^*)) = \sum_\alpha\sum_\beta\lambda_\alpha\lambda_\beta C(\bdsm{x}_\alpha - \bdsm{x}_\beta) - 2\sum_\alpha\lambda_\alpha C(\bdsm{x}_\alpha - \bdsm{x}^*) + C(\bdsm{0}),\]
subject to the unbiasedness constraint $\sum_\alpha\lambda_\alpha = 1$. The formulation under the variogram stationarity is similar.
The important takeaway from this model framework is that the second-order stationarity assumption makes the selection of the $\lambda_\alpha$ entirely dependent on the covariance or variogram, which is assumed to be a function of the spatial deviation between observations. 

The key to developing interval-valued kriging is to define a proper second-order structure for the interval-valued random function $[Z(\bdsm{x})]=[Z^L(\bdsm{x}), Z^U(\bdsm{x})]$. Diamond's approach used an interval-valued covariance as
\begin{equation}\label{eqn:diamond-cov}
\left[C(\bdsm{x}, \bdsm{h})\right] = E\left[Z(\bdsm{x})Z(\bdsm{x} + \bdsm{h})\right] - E\left[Z(\bdsm{x})\right]E\left[Z(\bdsm{x}+\bdsm{h})\right].
\end{equation}
This definition is only conceptual, because there is no inverse element in the space $\mathcal{K}_{\mathcal{C}}(\mathbb{R})$ and thus the subtraction of intervals is not defined. In addition, it requires the multiplication of intervals, which is complicated in general. For simplification purposes, Diamond restricted considerations to positive intervals, i.e., intervals that contain only positive numbers. Under this restriction, the multiplication is seen to be 
\[
\left[Z(\bdsm{x} + \bdsm{h})\right]\left[Z(\bdsm{x})\right] = \left[Z(\bdsm{x} + \bdsm{h})^LZ(\bdsm{x})^L, Z(\bdsm{x} + \bdsm{h})^UZ(\bdsm{x})^U\right].
\]
Diamond's notion of second-order stationarity was subsequently defined only for positive intervals with conditions:
\begin{enumerate}
\item $E\left[Z(\bdsm{x})\right]=[m]$ exists and is independent of $\bdsm{x}$;
\item $E\left[Z(\bdsm{x})Z(\bdsm{x} + \bdsm{h})\right]=\left[C(\bdsm{x}, \bdsm{h})\right] + \left[m\right]^2$ exists and is independent of $\bdsm{x}$, assuming $\left[C(\bdsm{x}, \bdsm{h})\right]$ is a positive interval. 
\end{enumerate}

There are several mathematical difficulties in this framework. First, as mentioned above, there is no well-defined subtraction operation for intervals and therefore the interval-valued covariance $\left[C(\bdsm{x}, \bdsm{h})\right]$ cannot be determined by (\ref{eqn:diamond-cov}). Notice how the covariance stationarity (condition 2) is stated indirectly by $E\left[Z(\bdsm{x})Z(\bdsm{x} + \bdsm{h})\right]$, instead of by the covariance $\left[C(\bdsm{x}, \bdsm{h})\right]$ itself.  While one can theoretically impose a covariance structure by assumption, in practice it is not obvious how the covariance can be estimated from the data, which limits its applicability. The second difficulty is with regard to the mathematical coherence of the variance and covariance, that is, the covariance of two identical quantities should be the same as the variance. However, it can be seen that 
$\text{Var}\left[Z(\bdsm{x})\right]\ne C(\bdsm{x}, \bdsm{0})$. In fact, the former is real-valued and the latter is an interval. Lastly, to ensure the non-negativity of the prediction variance, all of the interpolation weights are assumed to be non-negative, which makes the model even more restrictive. 

According to the recent development of set-valued statistics (e.g., \cite{Korner97, Korner98}), the covariance of random intervals, and in general random sets, should be defined as real-valued. This could be the potential solution to the aforementioned problems in Diamond's formulation. Motivated by this, we propose to re-construct the second-order structure based on the random sets theory to modify Diamond's interval-valued kriging into a more rigorous and more computationally feasible method. To this, the notion of variance of a random set \citep{Lyashenko82, Nather97, Korner95, Korner97} plays the key role. Given a metric $\rho$ in the space $\mathcal{K}$, the variance of a random compact set $A$ is defined as $\text{Var}_{\rho}(A)=E\rho^2[A, E(A)]$. Now if we restrict to $\mathcal{K}_{\mathcal{C}}$, according to the embedding, a random compact convex set $X$ can be represented by its support function $s_X$ and the space $\mathcal{K}_{\mathcal{C}}$ is equipped with an $L_2$ metric $\rho_2$. Considering $<\cdot, \cdot>$ as the inner product in the Hilbert space $L_2(S^{d-1})$, the variance is defined as
\[
  \text{Var}(X)=E\left\|s_{X}-s_{E(X)}\right\|_2^2=E\int_{S^{d-1}}[s_{X}-s_{E(X)}]^2\mu d(u)=E<s_{X}-s_{E(X)}, s_{X}-s_{E(X)}>. 
\]
This leads to the natural extension to the covariance function for $X, Y\in\mathcal{K}_{\mathcal{C}}(\mathbb{R}^d)$ as
\[
  \text{Cov}(X, Y)=E<s_X-s_{E(X)}, s_Y-s_{E(Y)}>=E\int_{S^{d-1}}[s_X-s_{E(X)}][s_Y-s_{E(Y)}]\mu d(u). 
\]
Such a definition of covariance has been shown to be very favorable for statistical analysis \citep{Korner95, Korner97}. 
Consider random intervals $[X],[Y]\in\mathcal{K}_\mathcal{C}(\mathbb{R})$ and the general metric $\rho_K$. The variance is seen to be
\begin{eqnarray}
  \text{Var}([X])&=&E\left\{\rho^2_K([X], E([X]))\right\}\nonumber\\
  &=&E\big[A_{11}\left(X^C-E(X^C)\right)^2+A_{22}\left(X^R-E(X^R)\right)^2 \nonumber \\
  &&+2A_{12}\left(X^C-E(X^C)\right)\left(X^R-E(X^R)\right)\big]\nonumber\\
  &=&A_{11}\text{Var}(X^C)+A_{22}\text{Var}(X^R)+2A_{12}\text{Cov}(X^C, X^R),\label{def:int_var}
\end{eqnarray}
The covariance is a little more complex and relies on writing the inner product associated with the $\rho_K$ metric as 
\begin{eqnarray*}
  <s_X-s_{E(X)}, s_Y-s_{E(Y)}>
  &=&B_{11}\left(X^C-E(X^C)\right)\left(Y^C-E(Y^C)\right)\\
  && +B_{22}\left(X^R-E(X^R)\right)\left(Y^R-E(Y^R)\right)\\
  &&+B_{12}\left(X^C-E(X^C)\right)\left(Y^R-E(Y^R)\right)\\
  &&+B_{21}\left(X^R-E(X^R)\right)\left(Y^C-E(Y^C)\right),
\end{eqnarray*}
where $B$ is a symmetric positive definite matrix uniquely determined by $K$. The covariance is consequently defined as
\begin{eqnarray*}
  \text{Cov}([X],[Y])&=&E\left\{<s_X-s_{E(X)}, s_Y-s_{E(Y)}>\right\}\\
  &=&B_{11}\text{Cov}(X^C, Y^C)+B_{22}\text{Cov}(X^R, Y^R) \\
  && +B_{12}\text{Cov}(X^C, X^R)+B_{21}\text{Cov}(X^R, X^C). 
\end{eqnarray*}

We are now ready to introduce our interval-valued kriging and the definition of stationarity. Recall that $[Z(\bdsm{x})]=[Z^L(\bdsm{x}), Z^U(\bdsm{x})]$ denotes the interval-valued random function, which can be alternatively represented by the center function $Z^C(\bdsm{x})$ and the radius function $Z^R(\bdsm{x})$. As in \cite{Diamond88}, our interval-valued kriging interpolator is defined as
\[\label{def:kriging}
  \widehat{[Z]}(\bdsm{x}^*)=\sum_{i=1}^{n}\lambda_i[Z](\bdsm{x}_i),
\]
according to the Minkowski addition and scalar multiplication. It can be expressed equivalently in the center-radius form as
\[
  \hat{Z}^C(\bdsm{x}^*)=\sum_{i=1}^{n}\lambda_iZ^C(\bdsm{x}_i),\ \ \ \ \hat{Z}^R(\bdsm{x}^*)=\sum_{i=1}^{n}|\lambda_i|Z^C(\bdsm{x}_i).
\]
Given the preceding discussion of the second-order structure of random intervals, the stationarity of $[Z(\bdsm{x})]$ is derived from a natural extension of the stationarity for point-valued random function. We formally state it in the following. 
\begin{definition}\label{def:stationarity}
The interval-valued random function $[Z(\cdot)]$ is second-order stationary if it satisfies
\begin{enumerate}
\item (Mean Stationarity) $E([Z(\bdsm{x})])=[m]$, for some fixed interval $[m]$ independent of $\bdsm{x}$, i.e. $E(Z^C(\bdsm{x}))=m^C$ and $E(Z^R(\bdsm{x}))=m^R\geq0$ independent of $\bdsm{x}$;
\item (Covariance Stationarity) Cov$\left([Z(\bdsm{x}+\bdsm{h})], [Z(\bdsm{x})]\right), \bdsm{h}\in\mathbb{R}^n$ is a function of $\bdsm{h}$ only, i.e., the four covariance functions 
\text{Cov}$\left(Z^I(\bdsm{x}+\bdsm{h}), Z^J(\bdsm{x})\right)=C^{I,J}(\bdsm{h})$, $I,J\in\left\{C,R\right\}$, are all independent of $\bdsm{x}$.
\end{enumerate}
\end{definition}
As a remark on the covariance stationary, $C^{R,C}$ is completely determined by $C^{C,R}$ in that $C^{R,C}(\bdsm{h})=C^{C,R}(-\bdsm{h})$, so only three covariance functions, i.e., $C^{C,C}(\bdsm{h})$, $C^{R,R}(\bdsm{h})$, and $C^{C,R}(\bdsm{h})$ are needed to define stationarity. Under the assumption of second-order stationarity, the prediction variance of the kriging estimator is calculated in the following Theorem \ref{thm:kriging_var}.

\begin{theorem}\label{thm:kriging_var}
Up to an additive constant, the prediction variance of the interval-valued kriging interpolator is equal to
\small{
\begin{eqnarray}
E\left[\rho^2_K\left([\hat{Z}(x^*)], [Z(x^*)]\right)\right] 
&=&A_{11}\left[\sum_i\sum_j\lambda_i\lambda_jC^{C,C}(x_i-x_j)-2\sum_i\lambda_iC^{C,C}(x_i-x^*)\right]\nonumber\\
&&+A_{22}\left[\sum_i\sum_j\left|\lambda_i\lambda_j\right|C^{R,R}(x_i-x_j)-2\sum_i\left|\lambda_i\right|C^{R,R}(x_i-x^*)\right]\nonumber\\
&&+2A_{12}\left[\sum_i\sum_j\lambda_i|\lambda_i|C^{C,R}(x_i-x_j)\right] \nonumber \\
&& -2A_{12}\left[\sum_i|\lambda_i|C^{C,R}(x^*-x_i)+\sum_i\lambda_iC^{C,R}(x_i-x^*)\right].\label{eqn:pred-var-cov}
\end{eqnarray}
}
\end{theorem}

\subsection{Simple Kriging (SK)}
Assume $E\left([Z(\bdsm{x})]\right)=[m]$ for a known fixed interval $[m]$. We can replace $[Z]$ by $[Z]-m^C$ (and add $m^C$ back after the model is fitted) so that the center function has a constant mean of zero. Then, $\hat{Z}^C(\bdsm{x}^*)$ is automatically unbiased and the unbiasedness of $\hat{Z}^R(\bdsm{x}^*)$ implies $\sum_{i=1}^{n}|\lambda_i|=1$. 

Hence, the interval-valued SK estimator is defined as the minimizer of the prediction variance under the unbiasedness constraint, i.e., 
\begin{equation}\label{def:sk}
  [\hat{Z}^{SK}(x^*)]=\arg\min E\left[\rho^2_K\left([\hat{Z}(x^*)], [Z(x^*)]\right)\right],\ \ \text{subject to}\ \ \sum_{i=1}^{n}|\lambda_i|=1.
\end{equation}

\subsection{Ordinary Kriging (OK)}
OK still assumes that $E\left([Z(\bdsm{x})]\right)=[m]$, but the interval-valued mean $[m]$ is unknown. Thus, we can no longer demean the center and instead have to impose the additional condition $\sum_{i=1}^{n}\lambda_i=1$ to ensure that the center prediction is unbiased. This together with the condition $\sum_{i=1}^{n}|\lambda_i|=1$ implies that the weights need to be all non-negative. Therefore, the interval-valued OK estimator is defined as 
 \begin{eqnarray}\label{def:ok}
  [\hat{Z}^{OK}(x^*)]&=&\arg\min E\left[\rho^2_K\left([\hat{Z}(x^*)], [Z(x^*)]\right)\right],\nonumber \\ 
  && \text{subject to}\ \sum_{i=1}^{n}\lambda_i=1,\ \ \lambda_i\geq 0,\ \ i=1,\cdots, n. 
\end{eqnarray}

\subsection{The Variogram}
In a slightly different situation, instead of assuming $Z(\bdsm{x})$ is stationary, it is assumed that the increment $Z(\bdsm{x}+\bdsm{h})-Z(\bdsm{x})$ is stationary. A point-valued random function that is second-order increment stationary is defined such that the first two moments of the increment are both independent of $\bdsm{x}$, i.e., 
\begin{eqnarray*}
  &&E\left[Z(\bdsm{x}+\bdsm{h})-Z(\bdsm{x})\right]=m\left(\bdsm{h}\right);\\
  &&\text{Var}\left[Z(\bdsm{x}+\bdsm{h})-Z(\bdsm{x})\right]=2\gamma\left(\bdsm{h}\right), 
\end{eqnarray*}
where $\gamma(\bdsm{h})$ is called the semi-variogram. Usually the function $m(\bdsm{h})$ is assumed to be constantly zero, namely, $Z(\bdsm{x})$ has a constant mean. Increment stationarity is a slightly weaker condition than stationarity, mainly because it allows the variance of $Z(\cdot)$ to be infinite. For interval-valued random function $[Z(\bdsm{x})]$, it is difficult to define ``increment'' as in the point-valued case, because there is no inverse element of addition in the space $\mathcal{K}_{\mathcal{C}}(\mathbb{R})$. Nevertheless, the assumptions for increment stationarity can be equivalently specified for interval-valued process through an interval-valued drift function and a real-valued semi-variogram as follows:
\begin{definition}
The interval-valued random function $[Z(\cdot)]$ is increment stationary if it satisfies
\begin{enumerate}
\item (Mean Stationarity) $E\left([Z(\bdsm{x})]\right)=[m]$, for some fixed interval $[m]$ independent of $\bdsm{x}$;
\item (Variogram Stationarity) $E\rho_K^2\left([Z(\bdsm{x}+\bdsm{h})], [Z(\bdsm{x})]\right)=2\gamma(\bdsm{h})$, $h\in\mathbb{R}^n$, independent of $\bdsm{x}$. 
\end{enumerate}
\end{definition}

According to the definition of variance (\ref{def:int_var}), the semi-variogram $\gamma(\bdsm{h})$ breaks down into the center, radius, and center-radius semi-variograms as 
\begin{eqnarray*}
  \gamma(\bdsm{h})=A_{11}\gamma^C(\bdsm{h})+A_{22}\gamma^R(\bdsm{h})+2A_{12}\gamma^{C,R}(\bdsm{h}),
\end{eqnarray*}
where
\begin{eqnarray*}
  &&\gamma^C(\bdsm{h})=(1/2)\text{Var}\left(Z^C(\bdsm{x}+\bdsm{h})-Z^C(\bdsm{x})\right),\\ 
  &&\gamma^R(\bdsm{h})=(1/2)\text{Var}\left(Z^R(\bdsm{x}+\bdsm{h})-Z^R(\bdsm{x})\right),\\
  &&\gamma^{C,R}(\bdsm{h})=(1/2)\text{Cov}\left(Z^C(\bdsm{x}+\bdsm{h})-Z^C(\bdsm{x}), Z^R(\bdsm{x}+\bdsm{h})-Z^R(\bdsm{x})\right). 
\end{eqnarray*} 
Thus, variogram stationarity means that all of the semi-variograms $\gamma^C$, $\gamma^R$, and $\gamma^{C,R}$, are independent of $\bdsm{x}$. In practice, their forms can be chosen by the corresponding sample estimates. If the covariance functions exist, they are related to the semi-variograms by the following equations:
\begin{eqnarray*}
  \gamma^C(\bdsm{h})&=&C^{C,C}(\bdsm{0})-C^{C,C}(\bdsm{h}),\label{eqn:vario-cov-c}\\
  \gamma^R(\bdsm{h})&=&C^{R,R}(\bdsm{0})-C^{R,R}(\bdsm{h}),\label{eqn:vario-cov-r}\\
  \gamma^{C,R}(\bdsm{h})&=&C^{C,R}(\bdsm{0})-(1/2)\left[C^{C,R}(\bdsm{h})+C^{R,C}(\bdsm{h})\right],\label{eqn:vario-cov-cr1}\\
  &=&C^{C,R}(\bdsm{0})-(1/2)\left[C^{C,R}(\bdsm{h})+C^{C,R}(-\bdsm{h})\right].\label{eqn:vario-cov-cr2}
\end{eqnarray*} 

\begin{theorem}\label{thm:kriging_var_vario}
Under the unbiasedness constraints $\sum_{i=1}^{n}\lambda_i=1$ and $\sum_{i=1}^{n}|\lambda_i|=1$, $i=1,\cdots, n$, the prediction variance is equal to
\begin{eqnarray*}
&&E\left[\rho^2_K\left([\hat{Z}(x^*)], [Z(x^*)]\right)\right]\nonumber\\
&=&A_{11}\left[-\sum_i\sum_j\lambda_i\lambda_j\gamma^C(x_i-x_j)+2\sum_i\lambda_i\gamma^C(x_i-x^*)\right]\nonumber\\
&+&A_{22}\left[-\sum_i\sum_j\lambda_i\lambda_j\gamma^R(x_i-x_j)+2\sum_i\lambda_i\gamma^C(x_i-x^*)\right]\nonumber\\
&+&2A_{12}\left[-\sum_{i\ne j}\lambda_i\lambda_j\gamma^{C,R}(x_i-x_j)+2\sum_i\lambda_i\gamma^{C,R}(x_i-x^*)\right].\label{eqn:pred-var-vario}
\end{eqnarray*}
\end{theorem}

\section{Numerical implementation}\label{numerics}
Implementation of the proposed SK and OK models amounts to minimizing the prediction variance subject to certain constraints. If the covariance functions exist, which implies that the variograms also exist, the prediction variance can be expressed either by (\ref{eqn:pred-var-cov}) or (\ref{eqn:pred-var-vario}). Otherwise, under a weaker assumption when the covariance functions do not exist but the variograms exist, the prediction variance is given by (\ref{eqn:pred-var-vario}). In either case, direct differentiation is impossible due to the involvement of $|\bdsm{\lambda}|$. In addition, the inequality constraints associated with the OK in (\ref{def:ok}) are a form of the Karush-Kuhn-Tucker conditions, for which an analytical solution usually does not exist. Considering all these, to implement the interval-valued kriging models, we propose a penalized approximate Newton-Raphson (PANR) algorithm that finds a numerical solution to the constrained optimization problem.   

\subsection{The penalty method for constraints}
Denote the prediction variance by 
\[
V(\bdsm{\lambda})= E\left[\rho^2_K\left([\hat{Z}(x^*)], [Z(x^*)]\right)\right].
\] 
Recall from (\ref{def:sk}) and (\ref{def:ok}) that SK minimizes $V(\bdsm{\lambda})$ subject to $\sum_{i=1}^{n}|\lambda_i|=1$ and OK subject to $\sum_{i=1}^{n}\lambda_i=1,\ \ \lambda_i\geq 0,\ \ i=1,\cdots, n$, respectively. We employ a penalty method to account for these optimization constraints. (See, e.g., \cite{Jensen03} for a review of algorithms for constrained optimization.) The idea is to approximate the constrained optimization problem by an unconstrained problem formulated as
\[\label{def:penalty}
  \arg\min \left\{V(\bdsm{\lambda}) + P(\bdsm{\lambda}, c)\right\},
\] 
where $P(\bdsm{\lambda}, c)$ is a continuous penalty function that equals to zero if and only if the constraints are satisfied and $c$ is a positive constant controlling the magnitude
of the penalty function. With an appropriately chosen penalty function, the solution of (\ref{def:penalty}) is approximately the same as the constrained minimizer of the original objective function $V(\bdsm{\lambda})$. Corresponding to the equality constraint $\sum_{i=1}^{n}|\lambda_i|=1$ for SK, the most natural penalty is the quadratic loss penalty 
\[\label{def:p-sk}
  P^{SK}(\bdsm{\lambda}, c)=(1/c)\left(1-\sum_{i=1}^{n}|\lambda_i|\right)^2.
\]
When $(1/c)$ is large enough, any violation of the constraint will result in a heavy cost from the penalty and thus minimizing the penalized objective function will yield a feasible solution. For OK, there are  an equality constraint $\sum_{i=1}^{n}\lambda_i=1$ and inequality constraints $\lambda_i\geq 0,\ \ i=1,\cdots, n$. To tackle such a problem, the most common strategy is to employ the logarithmic-quadratic loss function 
\[\label{def:p-ok}
  P^{OK}(\bdsm{\lambda}, c)=-c\sum_{i=1}^{n}\ln\left(\lambda_i\right) + (1/c)\left(1-\sum_{i=1}^{n}|\lambda_i|\right)^2,
\]
where the logarithmic terms take care of the inequality constraints. Similar to the pure quadratic loss penalty, small values of $c$ will lead to a solution within the feasible region. A simple straightforward strategy to implement the penalty method is known as the sequential unconstrained minimization technique (SUMT) \citep{Jensen03}. It starts with an initial value of the penalty parameter $c_0$ and iteratively updates it until the convergence criterion is satisfied. It was shown in \cite{Fiacco68} that for a sequence of monotonically increasing (or decreasing depending on the nature of the problem) $\left\{c_k\right\}$, the SUMT converges to the penalized objective within the feasible region. A small issue with this algorithm is that values of $c$ that are too large (or too small) will create ill-behaved surfaces for which gradient and Hessian calculations will be unstable. Therefore, slowly changing the values of $c$ balances the influence of the penalty term with the rest of the objective function and is the key to the success. 

For SK and OK, the penalty parameter $c$ needs to strictly decrease to a sufficiently small value to satisfy the unbiasedness constraint. OK has the additional stipulation that $\min(\bdsm{\lambda}^{(k)}) \ge 0$. This in mind, we implement the SUMT algorithm as follows: 
\begin{enumerate}
\item ({\it Initialization}) Set an initial for $c_0$ (for OK, we use $c_0=100$ and for SK, we use $c_0 = 0$) and determine an initial value $\bdsm{\lambda}^{(0)}$. 
\item ({\it Minimization}) Minimize $V(\bdsm{\lambda}) + P(\bdsm{\lambda}, c_k)$ to obtain $\bdsm{\lambda}^{(k+1)}$. Let $k=k+1$. 
\item ({\it Check Constraint}) Compute $p_k = 1 - \sum_{i=1}^{n}|\lambda^{(1)}_i|$. If $|p_k| < tolp$, where $tolp$ is a user defined tolerance for the constraint, end the iteration. Otherwise, go to Step 4. For OK, suspend the algorithm and return an error if $\min(\bdsm{\lambda}^{(k)}) < 0$.
\item ({\it Update Parameter}) Set $c_{k+1} = \begin{cases} c_{k}/\eta & \text{SK} \\ \eta c_{k} & \text{OK} \end{cases}$, with $\eta \in (0, 1)$ being a user defined parameter, and repeat Step $2$ and $3$.
\end{enumerate}

For both SK and OK, the algorithm is terminated if the minimization (Step 2) fails for the final value of $c_k$ after a user-specified maximum number of iterations ($maxq$), or the constraints are not satisfied within the maximum number of penalty iterations ($maxp$).  

\subsection{Approximation of \texorpdfstring{$|\bdsm{\lambda}|$}{lambda}}
The key step of the aforementioned SUMT algorithm is the minimization of the penalized prediction variance 
\begin{equation}\label{def:pvar}
Q(\bdsm{\lambda}) = V(\bdsm{\lambda}) + P(\bdsm{\lambda}, c).
\end{equation}
Numerically, this can be carried out by the PANR algorithm. In order to guarantee convergence, the objective function must be second-order continuously differentiable. However, the absolute value $|\lambda|$ in the prediction variance is not differentiable at $\lambda=0$. To address this issue, we propose to approximate the absolute value by a local quadratic function. Consider the Taylor expansion of $|\lambda|$ at $\lambda_0\ne 0$: 
\[
  |\lambda|=|\lambda_0|+|\lambda|'(\lambda_0)\left(\lambda-\lambda_0\right)+o\left(|\lambda-\lambda_0|^2\right),\ \ \lambda\approx\lambda_0. 
\] 
Replacing the derivative $|\lambda|'(\lambda_0)=sgn(\lambda_0)$ by $(1/c)$, $|\lambda|$ is approximated by a quadratic function as 
\begin{equation}\label{eqn:approx}
  |\lambda|\approx|\lambda_0|+\left(\lambda-\lambda_0\right)(1/c), \ \ \lambda_0\ne 0. 
\end{equation}
Under this approximation, the gradient $\bdsm{G}$ and Hessian $\bdsm{H}$ of $Q(\bdsm{\lambda})$ are defined as 
\[
  \bdsm{G} = \nabla Q(\bdsm{\lambda}) \qquad \bdsm{H} = \nabla^2 Q(\bdsm{\lambda}).
\]
and the iteration of the PANR algorithm is given by
\[
  \bdsm{\lambda}^{(m+1)} = \bdsm{\lambda}^{(m)} - \bdsm{H}^{-1}\left(\bdsm{\lambda}^{(m)}\right)*\bdsm{G}\left(\bdsm{\lambda}^{(m)}\right).
\]
where $\bdsm{\lambda}^{(m)}$ is the value of $\bdsm{\lambda}$ from the $m^{th}$ iteration. In the Theorem \ref{thm:Gradient-Hessian} below, the gradient and Hessian of $Q(\bdsm{\lambda})$ are explicitly calculated for both SK and OK. 
\begin{theorem}\label{thm:Gradient-Hessian}
Consider minimizing the penalized prediction variance $Q(\bdsm{\lambda})$ defined in (\ref{def:penalty}) and (\ref{def:pvar}) using Newton-Raphson algorithm. 
Let $|\bdsm{\lambda}|$ be approximated by (\ref{eqn:approx}), where $\bdsm{\lambda}_0$ is an approximated value of $\bdsm{\lambda}$ such that $\bdsm{\lambda}_0\ne\bdsm{0}$. Define \[f^{R,R}(\bdsm{\lambda}, \lambda_k) = \left[\sum_{i}\left|\lambda_i\right|\left[C^{R,R}(\bdsm{x}_i - \bdsm{x}_k) + C^{R,R}(\bdsm{x}_k - \bdsm{x}_i)\right] - 2C^{R,R}(\bdsm{x}_k - \bdsm{x}^*)\right].\] Then the gradient $\bdsm{G}$ and Hessian $\bdsm{H}$ of $Q(\bdsm{\lambda})$ for SK are 
\begin{eqnarray*}
\bdsm{G}_k &=& A_{11} \left[\sum_i\lambda_i\left[C^{C,C}(\bdsm{x}_i - \bdsm{x}_k) + C^{C,C}(\bdsm{x}_k - \bdsm{x}_i)\right] - 2C^{C,C}(\bdsm{x}_k - \bdsm{x}^*)\right]  \\
&& + A_{22} \left[(\lambda_k/|\lambda_{k0}|)f^{R,R}(\bdsm{\lambda}, \lambda_k)\right]\\
&& + 2A_{12}\left[ \sum_i \left[|\lambda_i|C^{C,R}(\bdsm{x}_k - \bdsm{x}_i) +  (\lambda_k/|\lambda_{k0}|)\lambda_iC^{C,R}(\bdsm{x}_i - \bdsm{x}_k)\right]\right] \\
&&- 2A_{12}\left[(\lambda_k/|\lambda_{k0}|)C^{R,C}(\bdsm{x}_k - \bdsm{x}^*) + C^{C,R}(\bdsm{x}_k - \bdsm{x}^*)\right] \\
&&- (2c\lambda_k/|\lambda_{k0}|)\left(1-\sum_i|\lambda_i|\right) \ \ \ \  k=1,\cdots, n;
\end{eqnarray*}
\begin{eqnarray*}
\bdsm{H}_{k,l} &=& A_{11}\left[ C^{C,C}(\bdsm{x}_l - \bdsm{x}_k) + C^{C,C}(\bdsm{x}_k - \bdsm{x}_l)\right] \\ 
 &&+ 
A_{22}\left[(\lambda_k\lambda_l/|\lambda_{k0}\lambda_{l0}|)\left[C^{R,R}(\bdsm{x}_l - \bdsm{x}_k) + C^{R,R}(\bdsm{x}_k - \bdsm{x}_l)\right] + (I_{\{k=l\}}(\bdsm{\lambda})/|\lambda_{k0}|)f^{R,R}\left(\bdsm{\lambda}, \lambda_k\right)\right] \\
&&+ 2A_{12}\left[(\lambda_l/|\lambda_{l0}|)C^{C,R}(\bdsm{x}_k - \bdsm{x}_l) + (\lambda_k/|\lambda_{k0}|)C^{C,R}(\bdsm{x}_l - \bdsm{x}_k)\right] \\
&&+(2A_{12}I_{\{k=l\}}(\bdsm{\lambda})/|\lambda_{k0}|)\left[\sum_i\lambda_iC^{C,R}(\bdsm{x}_i - \bdsm{x}_k) - C^{R,C}(\bdsm{x}_k - \bdsm{x}^*)\right] \\
 &&+ (2c\lambda_k\lambda_l|\lambda_{k0}||\lambda_{l0}|) - I_{\{k=l\}}(\bdsm{\lambda})\left[(2c/|\lambda_{k0}|)\left(1-\sum_i|\lambda_i|\right)\right] 
\ \ \ \  k,l=1,\cdots, n.
\end{eqnarray*}

For OK, 
\begin{eqnarray*}
\bdsm{G}_k &=& A_{11}\left[\sum_i\lambda_i\left[C^{C,C}(\bdsm{x}_i - \bdsm{x}_k) + C^{C,C}(\bdsm{x}_k - \bdsm{x}_i)\right] - 2C^{C,C}(\bdsm{x}_k - \bdsm{x}^*)\right] \\
&&+ A_{22}\left[\sum_{i}\lambda_i\left[C^{R,R}(\bdsm{x}_i - \bdsm{x}_k) + C^{R,R}(\bdsm{x}_k - \bdsm{x}_i)\right] - 2C^{R,R}(\bdsm{x}_k - \bdsm{x}^*)\right] \\
&&+ 2A_{12}\left[\sum_i \lambda_i\left[C^{C,R}(\bdsm{x}_k - \bdsm{x}_i) + C^{C,R}(\bdsm{x}_i - \bdsm{x}_k)\right]\right] \\
&& - 2A_{12}\left[C^{R,C}(\bdsm{x}_k - \bdsm{x}^*) + C^{C,R}(\bdsm{x}_k - \bdsm{x}^*)\right] \\
&&- (2/c)\left(1-\sum_i\lambda_i\right) - (c/\lambda_k)\ \ \ \  k=1,\cdots,n;   \\
\\
\bdsm{H}_{k,l} &=& A_{11}\left[C^{C,C}(\bdsm{x}_l - \bdsm{x}_k) + C^{C,C}(\bdsm{x}_k - \bdsm{x}_l)  \right]   \\
&& +A_{22}\left[C^{R,R}(\bdsm{x}_l - \bdsm{x}_k) + C^{R,R}(\bdsm{x}_k - \bdsm{x}_l) \right] \\
&&+ A_{12}\left[C^{C,R}(\bdsm{x}_k - \bdsm{x}_l) + C^{C,R}(\bdsm{x}_l - \bdsm{x}_k) \right] \\
&& +(2/c) + (c/\lambda_k^2)I_{\{k=l\}}(\bdsm{\lambda})
\ \ \ \  k,l=1,\cdots, n.
\end{eqnarray*}
\end{theorem} 

\subsection{Adjustments for effective zero weights\label{0valAdj}}
The assumption for approximation (\ref{eqn:approx}) to work is that $\lambda_{i0}\ne 0$, $(i=1,\cdots,n)$. Thus, zero estimates are not allowed. Also, the calculation of $\ln\left(\lambda_i\right)$ gets very unstable when $\lambda_i$ is close to zero.  To guard against zero estimates, a natural strategy is to set small values to zero and exclude them from the next iteration. Such a strategy reduces the dimension of the gradient and Hessian and in some instances greatly speeds the computation. However, it suffers from a big drawback that once a weight parameter is set to zero, it remains zero in the ensuing iterations. This is potentially problematic, for example, in the event that some parameters are close to zero initially but would tend away from zero as the penalty iterates. To avoid this problem, instead of sub-setting the parameter vector at each iteration, we propose adjustments to the penalties, which still allow the $\lambda_i$'s to approach zero using the tolerance criteria described previously, yet initially prevents movement to zero. 

For SK, we simply add a barrier function that prohibits the $\lambda_i$'s from approaching zero. This leads to the following new penalty:
\begin{equation}\label{def:p-sk-adj}
\tilde{P}^{SK}(\bdsm{\lambda}, c) = -(c/n^2)\sum_i\ln\left(\lambda_i^2\right) + \left(\left(1-\sum_i|\lambda_i|\right)^2/c\right).
\end{equation}
Dividing the barrier penalty parameter by $n^2$ ensures a balance between the penalty terms so that one does not dominate the other. Now that the estimates are guarded against zero, the approximation of the absolute value function is no longer needed. The derviation of the gradient and Hessian in this case proceeds identically as in Theorem \ref{thm:Gradient-Hessian} using 
\[
  (d/d\lambda)|\lambda| = \sgn(\lambda),\ \ \ \ 
  (d^2/d\lambda^2)|\lambda| = 0.
\] 

For OK, in the current version of the penalty, the cost of reaching the zero-valued boundary is too high relative to the equality constraint. As a result, the algorithm tends to produce a non-zero weight for each lambda and predictions very close to the global average. Therefore, we add a small tolerance to the logarithmic penalty, which shifts the boundary of the penalty slightly below zero to allow $\lambda_i$ to reach zero without leaving the feasible region. In the mean time, any value that effectively dips below zero will be set equal to zero before the next iteration.  In addition, we divide the weight of the quadratic penalty by $n$ to ensure that growth of the equality constraint (as $c$ decreases) does not outpace the corresponding decrease in the $n$ inequality constraints. With these adjustments, the new penalty is given as
\[
\tilde{P}^{OK} = -c\sum_i\ln\left(\lambda_i + tolz\right) + \left(\left(1-\sum_i\lambda_i\right)^2/(c*n)\right).
\]

\section{Design snow load predictions for Utah}\label{body3:application}
Nearly all buildings in the USA are designed to strike a crucial balance
between safety and economy: a building must reasonably withstand the anthropogenic and
environmental forces induced upon it throughout its lifetime, yet use materials and designs
that are realistically affordable to the future occupants. One load of particular interest to mountainous states is the force induced by settled snow on the roof of the structure. Structures are typically built to withstand a design snow load, which is the force induced by snow expected to occur once every 50 years. Regional design snow load requirements are typically obtained by predicting design snow loads between measurement locations using an appropriate mapping technique.

The process of defining design snow loads is subject to several sources of imprecision including estimates of snow weight from snow depth, and estimates of future, extreme snow events using historical data. Current mapping techniques ignore this imprecision by treating the design snow loads estimated at measurement locations as exact observations. In this section, we illustrate how our interval-valued kriging models can preserve the imprecision in the input data due to the depth-to-load conversion process. We accomplish this by creating an interval-valued design snow load dataset with the same data used in the Utah Snow Load Study \citep{Bean2018-report}. Interval-valued predictions using our interval-valued data and kriging models are then compared to point-valued equivalents. Lastly, we discuss the notions of accuracy, precision, and variance in the interval-valued kriging framework and provide examples of each.  

\subsection{Defining the Interval-Valued Data}
Design snow loads often require estimating the weight of the snow pack from its depth. 
 Such estimations are notoriously difficult given the highly variable density of snow throughout the season. The need for appropriate snow density estimates has given rise to several models of varying complexity, many of which are discussed in \cite{Hill2019}. Consequently, design snow load requirements can vary substantially depending on the depth-to-load-conversion method used in the analysis. In this application, we focus on the current depth-to-load conversion techniques used in the most recent snow load studies of Colorado \citep{SEAC2016}, Idaho \citep{Sack2015-2}, and Utah \citep{Bean2018-report} with details provided in the appendix. Because it is unclear which of these three methods best characterizes the snowpack density at locations in and around Utah, we define interval-valued design snow loads $(\left[q^*_s(\bdsm{x}_\alpha)\right])$ through the following process. 

\begin{enumerate}
\item {\it Define a set of estimated annual maximum snow loads using the various depth-to-load conversion methods.}

Figure \ref{fig:sweCurves} compares depth-to-load prediction methods for various snow depths on different days of the year. Notice that each curve estimates a different load for a given snow depth. For each depth-to-load conversion method, we estimate and collect a set of annual maximum snow loads. Each set of annual maximum loads will be independently considered in the distribution fitting process of Step 2. 

\begin{figure}
\centering
\includegraphics[width=5in]{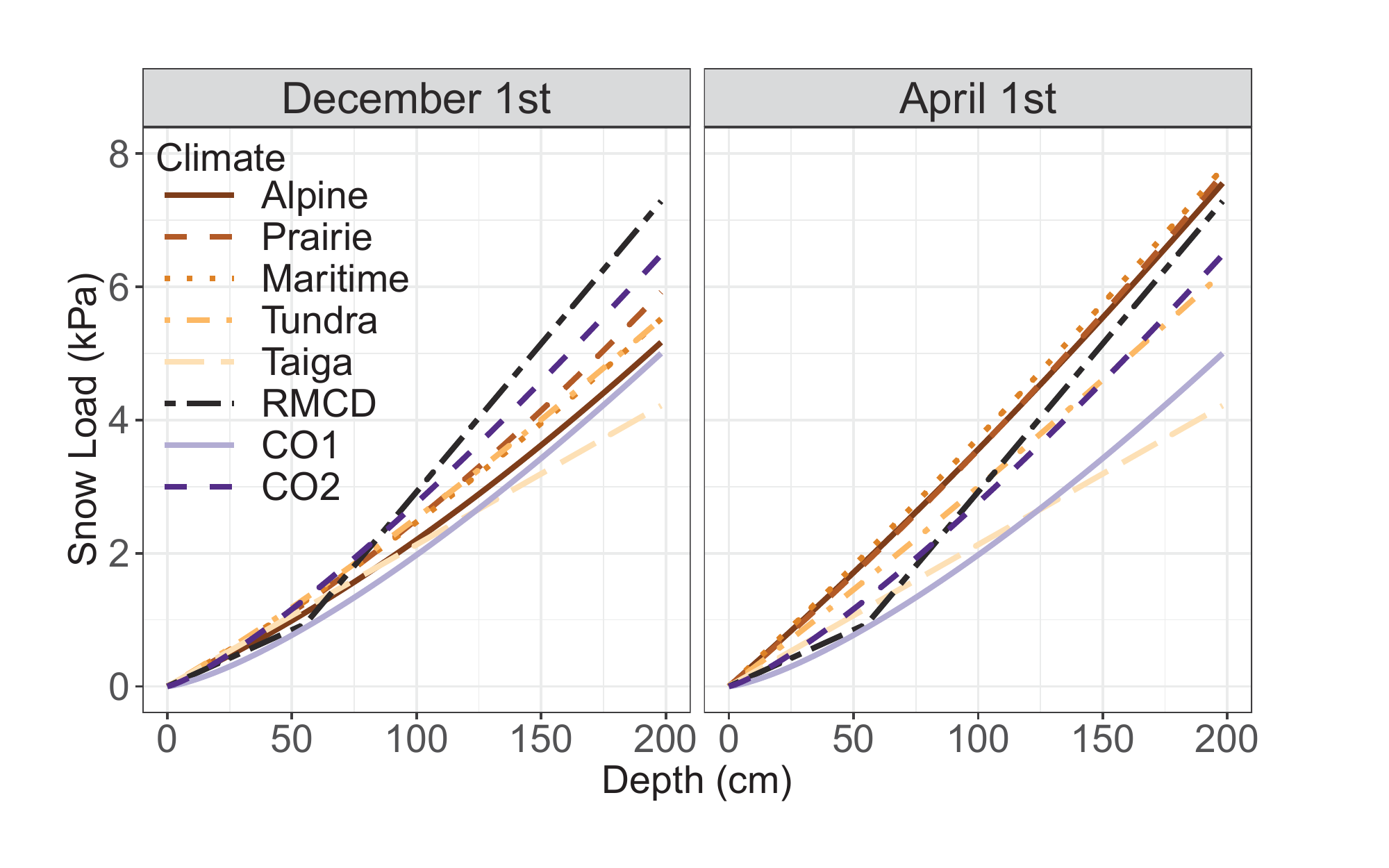}
\caption{Comparisons of the different depth-to-load conversion methods for various depths and days of the year.}
\label{fig:sweCurves}
\end{figure}
 
\item {\it Fit log-normal distributions to the annual maximum snow loads.}
The American Society of Civil Engineers (ASCE) conventionally assumes that annual maximum snow loads at a given location follow a log-normal distribution with the design snow load defined as the $98^\text{th}$ percentile of this distribution \citep{ASCE2017}. The Utah report follows this precedent \citep{Bean2018-report}, although other distributions have been used in other western states \citep{Sack2015}. \cite{Bean2019-cr} discusses the shortcomings of the distribution fitting approach to estimating design snow loads. Nevertheless, for the purpose of comparison, we likewise assume that any set of annual maximums resulting from Step 1 can be appropriately modeled by a log-normal distribution. 

Under this assumption, we independently fit log-normal distributions to each set of annual maximums produced by the various depth-to-load conversion methods. The 98th percentile from each of these distributions is defined as the estimated design ground snow load resulting from that particular depth-to-load conversion method.  

\item {\it Create a design snow load interval using the set of design snow load estimates.}

Step 2 creates a set of design ground snow load estimates $\bdsm{q}^*_s(\bdsm{x}_\alpha)$ resulting from each depth-to-load conversion method. We assume that the best design snow load estimate is contained within the range of these estimates. We therefore define the interval-valued design ground snow loads as $\left[q^*_s(\bdsm{x}_\alpha)\right] = \left[\text{min}(\bdsm{q}^*_s(\bdsm{x}_\alpha)), \text{max}(\bdsm{q}^*_s(\bdsm{x}_\alpha))\right]$. For locations that measure snow load directly, this process results in intervals of length zero. 
\end{enumerate}

\subsection{Analyses and results}
\begin{figure}
\centering
\includegraphics[width=0.5\textwidth]{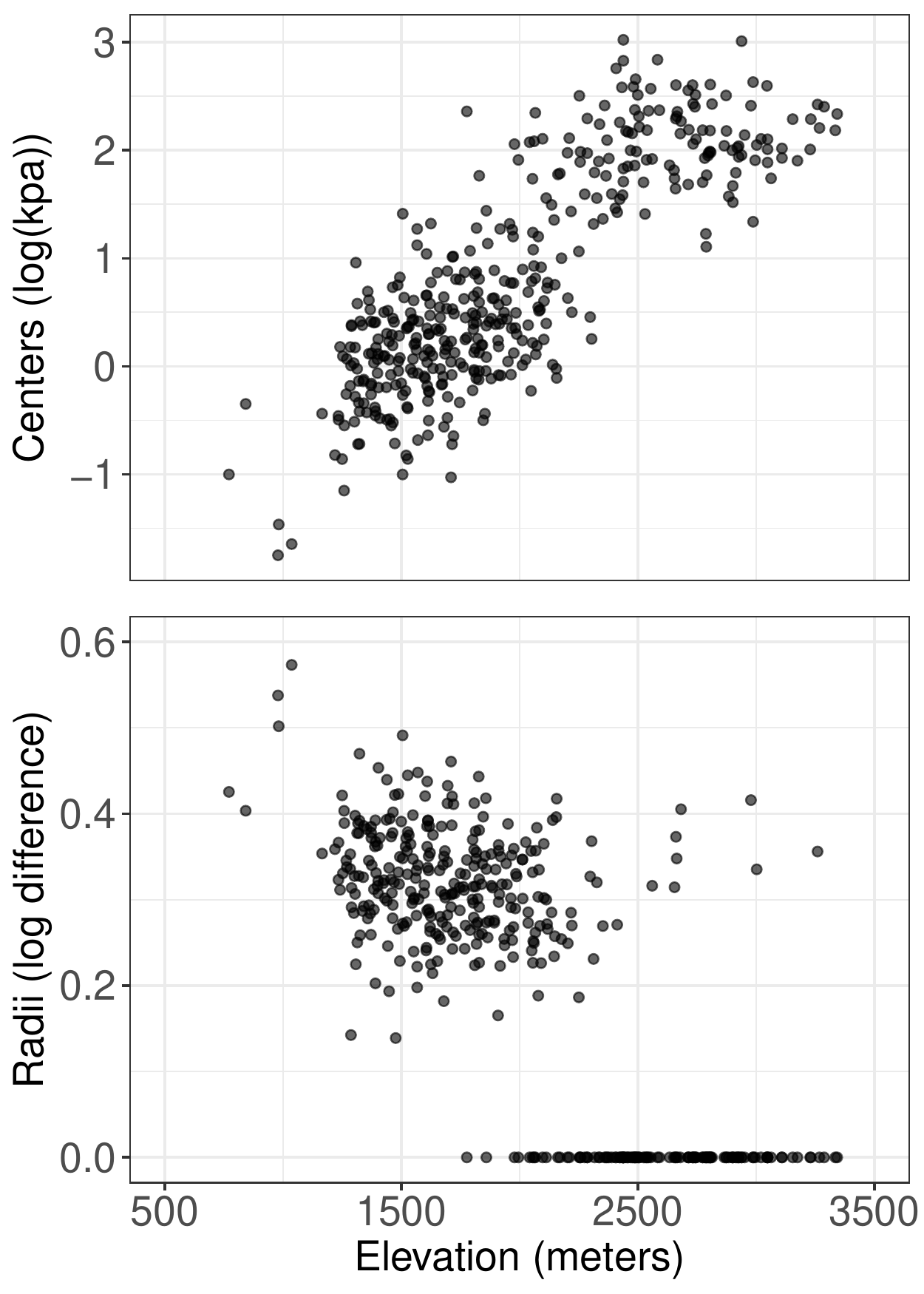}
\caption{Plots of interval centers and radii across elevation, emphasizing the need to account for the effect of elevation before assuming stationarity.}
\label{fig:centerCompare}
\end{figure}

Denote the final design snow load intervals by $[q_s^L(\bdsm{x}), q_s^U(\bdsm{x})]$. All ensuing analyses use these intervals on the log-scale, i.e.  
\[[\log(q_s^L(\bdsm{x})), \log(q_s^U(\bdsm{x}))] = [l_s^C(\bdsm{x}) - l_s^R(\bdsm{x}), l_s^C(\bdsm{x}) + l_s^R(\bdsm{x})].\] To achieve stationarity, we need to remove the elevation effect in both interval center and radius, shown in Figure \ref{fig:centerCompare}, prior to input into the kriging model. The elevation trends for both cases are modeled as 
\begin{eqnarray}
\label{eq:logLinear}
l_s^{C}(\bdsm{x}) &=& \beta_0 + \beta_1A(\bdsm{x}) + R^C(\bdsm{x}), \\
\label{eq:radLinear}
l_s^{R}(\bdsm{x}) &=& R^R(\bdsm{x})/\log(A(\bdsm{x})). 
\end{eqnarray}
The scaling strategy used for the residual radii ensures that $\min(R^R(\bdsm{x})) \ge 0$. Note that the radii scaling does not fully remove the effect of elevation as zero-valued radii only occur at higher elevations. Nevertheless, this scaling is effective at eliminating the significance of the Pearson correlation ($\rho = -.20$, $\text{p-val} = .0005$) between elevation and radii for all non-zero radius observations ($\rho = -.06$, $\text{p-val} = .28$).  It is the residual intervals, $\left[R^C - R^R, R^C + R^R\right]$, that are used as input in interval-valued kriging. 

\begin{figure}
\centering
\includegraphics[width=\textwidth]{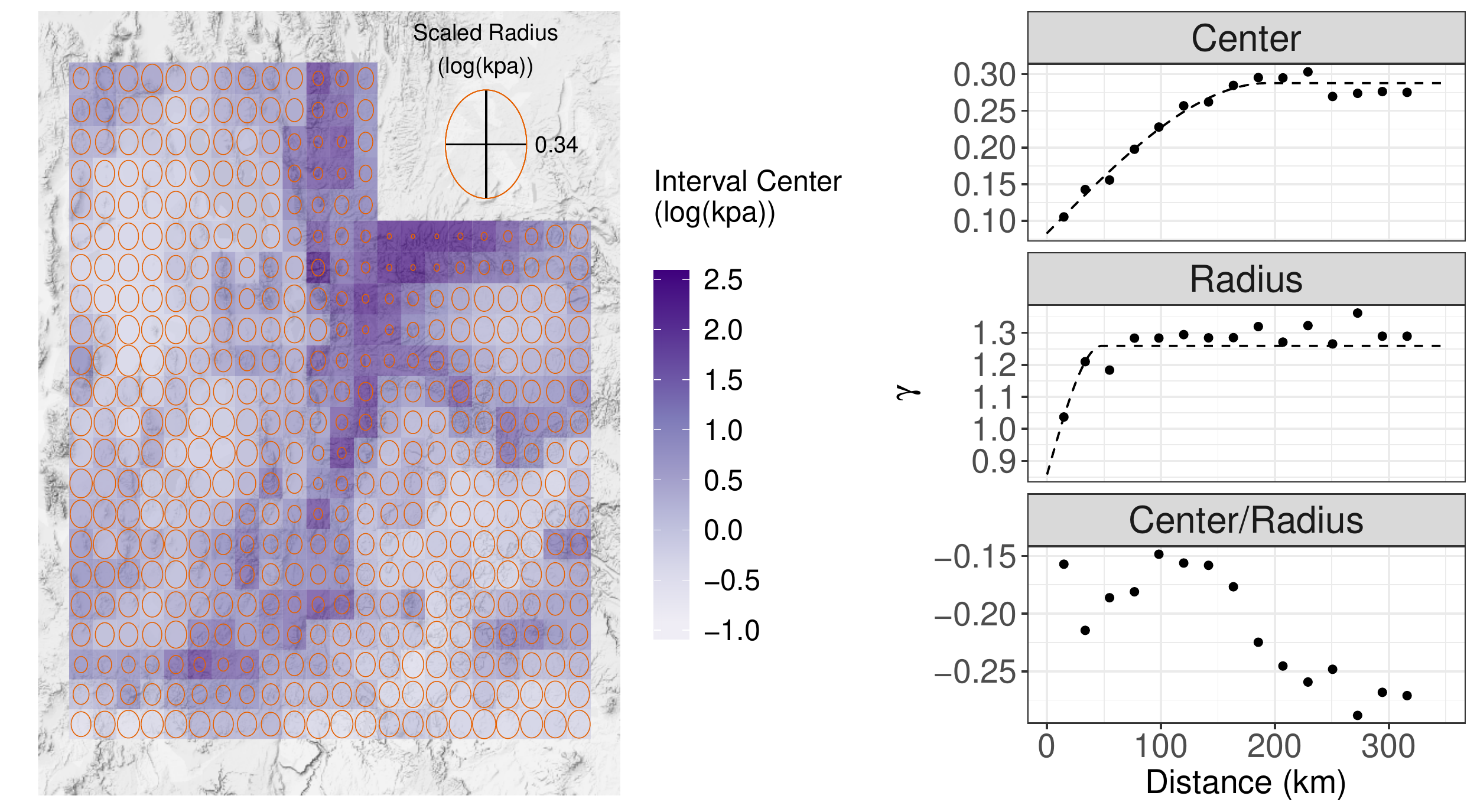}
\caption{(Left) An interval-valued design snow load prediction map for Utah. The darkness of the grid-cell indicates the interval center while the size of the circle within each cell represents the interval radius. Circles are scaled so that the maximum radius (indicated in the legend) is exactly circumscribed within the grid-cell. (Right) Plots of the empirical and theoretical variograms used in the numerical implementation of interval-valued simple kriging.}
\label{fig:utahVarios}
\end{figure}

Figure \ref{fig:utahVarios} shows the theoretical (lines) and empirical (points) variograms of the interval data alongside an interval-valued design snow load prediction map. The empirical variograms reveal spatial relationships between the centers and radii, but no obvious spatial relationship in the centers/radii interaction. For this reason, we choose to exclude the center/radius interaction from the $\rho_k$ metric calculation by setting $K = \begin{bmatrix} 0.5 & 0 \\ 0 & 0.5 \end{bmatrix}$. Such a strategy makes $\rho_k^2 = \rho_2^2$ and eliminates the need to fit a theoretical cross-variogram to the center/radius interaction. We selected spherical models for both the center and radius variograms (see \cite{Goovaerts1997} for a summary of commonly used variograms). Each model includes a nugget effect and parameters are selected using the weighted least squares fitting algorithm in the \texttt{gstat} package \citep{Bivand2013}. Their mathematical definitions are given as 
\begin{eqnarray*}
\gamma^C(||\bdsm{h}||) &=& \begin{cases} 0 & ||\bdsm{h}|| = 0, \\ 0.2\left[1.5(||\bdsm{h}||/194) - 0.5\left((||\bdsm{h}||/194)\right)^3\right] + 0.08 & 0 < ||\bdsm{h}|| \le 194,\\ 0.28 & ||\bdsm{h}|| > 194;\end{cases} \\
\gamma^R(||\bdsm{h}||) &=& \begin{cases} 0 & ||\bdsm{h}|| = 0, \\ 0.4\left[1.5(||\bdsm{h}||/48) - 0.5\left((||\bdsm{h}||/48)\right)^3\right] + 0.86 & 0 < ||\bdsm{h}|| \le 48,\\ 0.86 & ||\bdsm{h}|| > 48;\end{cases}
\end{eqnarray*}

We elected to use the SK model because $R^C(\cdot)$ has, both in theory and practice, a known mean of zero.
The resulting kriging predictions are input into (\ref{eq:logLinear}) and (\ref{eq:radLinear}) and exponentiated for final load predictions. An interval map of these final predictions shown in Figure \ref{fig:utahVarios} is intended for simultaneous visualizations of center and radius. The darkness of the grid indicates the interval center while the size of the circle within each grid represents the interval radius. The size of the circle is scaled so that the grid cell with the largest radius will have a circle exactly circumscribed within the cell.  

\begin{figure}
\centering
\includegraphics[width=.9\textwidth]{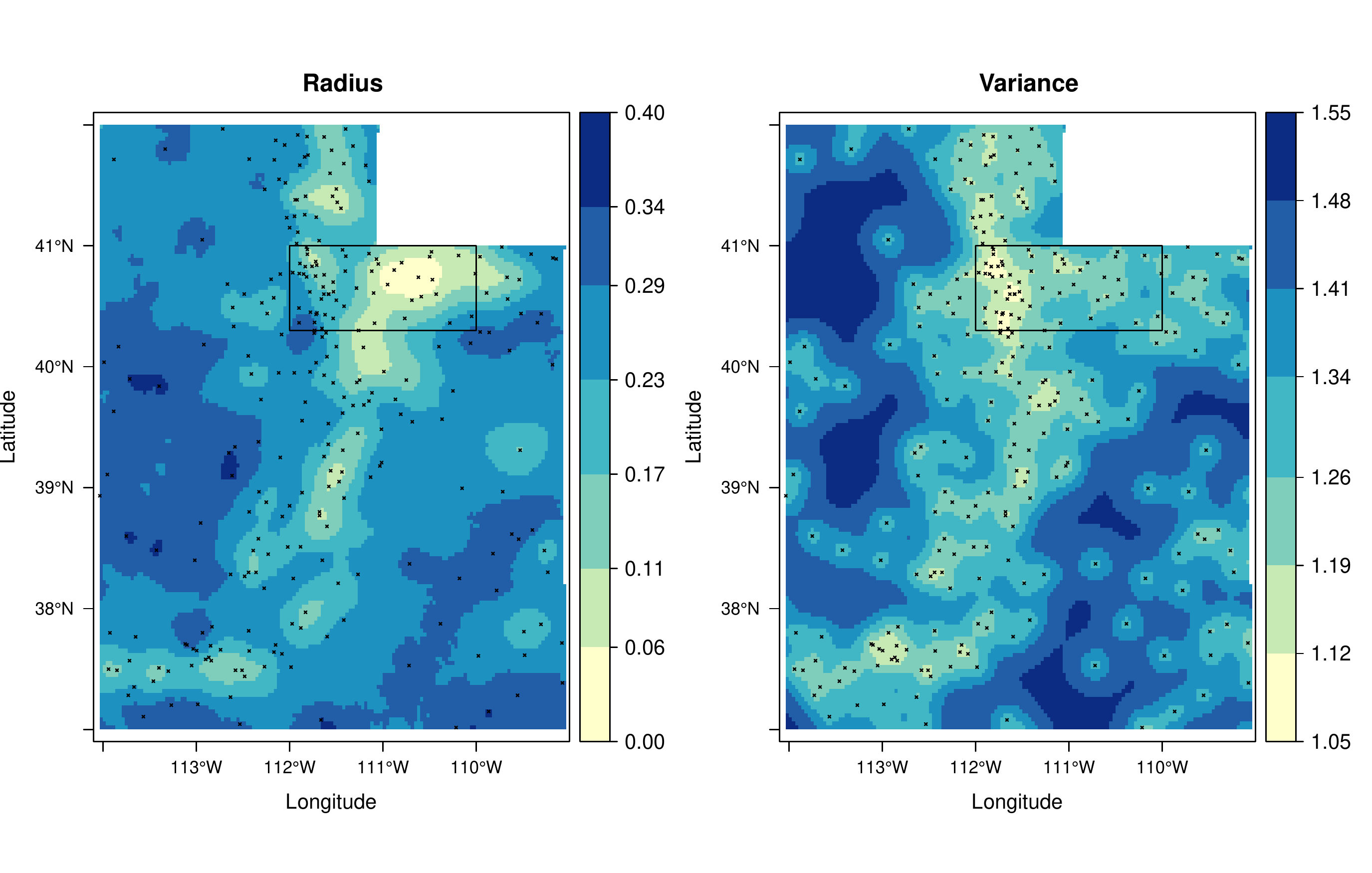}
\caption{Maps of the prediction variance and the predicted interval radii for the state of Utah. The black points denote Utah measurement locations.}
\label{fig:radiusVariance}
\end{figure}

It is important to distinguish the predicted interval radius from the prediction variance. Recall that the prediction variance defined in (\ref{eqn:pred-var-cov}) relies on the stationarity assumptions outlined in Definition \ref{def:stationarity}. One of these assumptions is that the covariances that comprise the prediction variance are solely a function of the location difference ($\bdsm{h}$). This essentially makes the kriging variance a measure of data quantity across space. In contrast, the interval radius measures the imprecision contributed by the surrounding measurements in prediction. Figure \ref{fig:radiusVariance} compares maps of kriging variance (defined in (\ref{eqn:pred-var-cov})) to the predicted interval radii.  The Pearson correlation between these two maps is significant (p-value < .0001) with a value of 0.55, but the two maps clearly measure different things. For example, the highlighted region in this figure shows that the lowest variance occurs in locations with the highest concentration of stations, while the lowest radii occurs in a region dominated by direct measurements of snow load (i.e. intervals of length 0). These separate measures emphasize the ability of interval-valued kriging to simultaneously account for the quantity \textit{and} quality of surrounding data in predictions. 

\subsection{Discussion}
\begin{figure}
\centering
\includegraphics[width=0.95\textwidth]{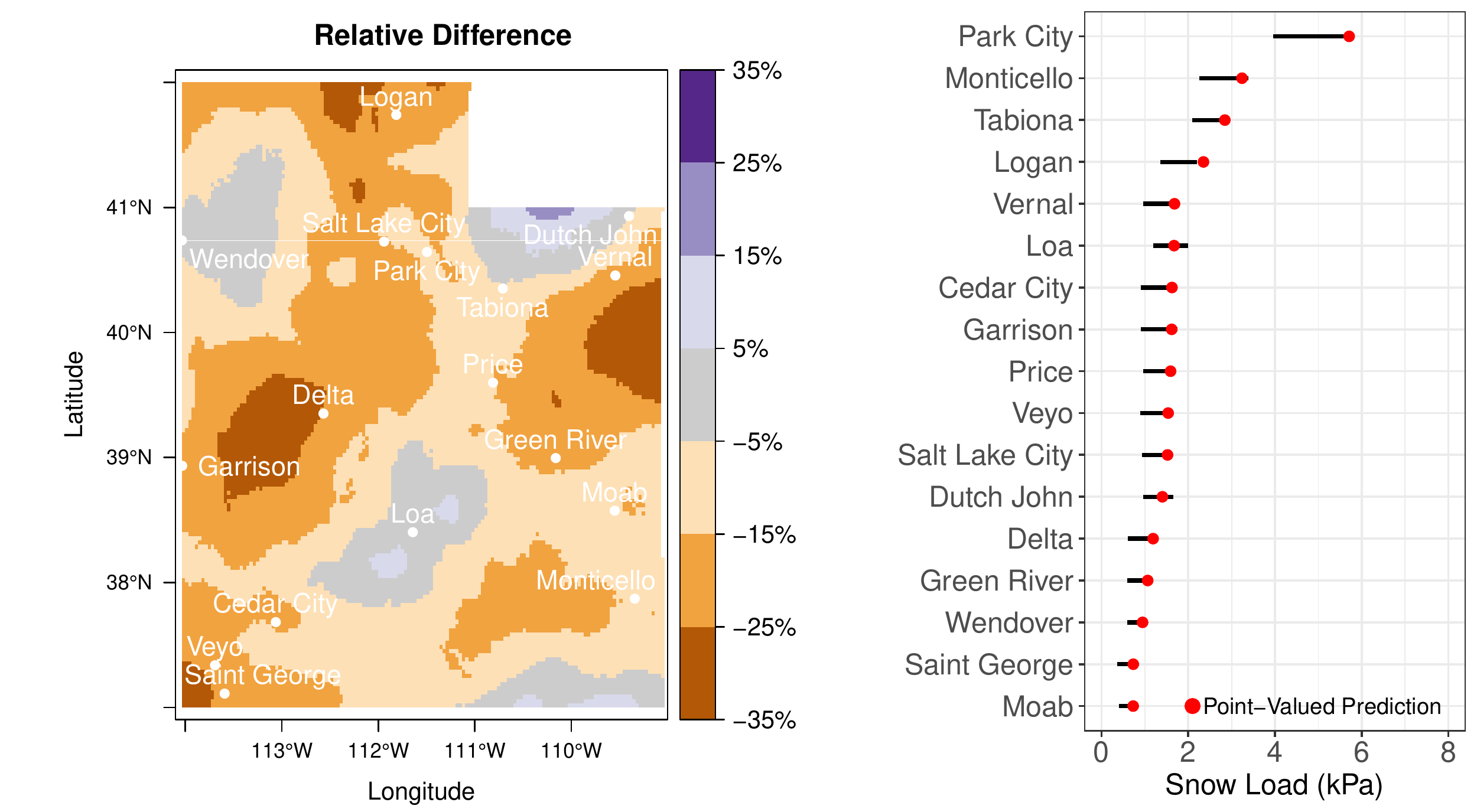}
\caption{(Left) Comparison of the relative difference between the predicted interval centers and the original point-valued predictions across space on the raw scale. Orange represents areas where the point predictions are higher than the interval centers and purple represents the contrary. (Right) Comparison of interval-valued simple kriging predictions to the original point-valued predictions at selected locations in Utah.}
\label{fig:poPredict}
\end{figure}

Figure \ref{fig:poPredict} compares the predicted intervals to the original point predictions using simple kriging with varying local means (SKLM) \citep{Goovaerts2000}. In the map, orange color represents areas where the point predictions are relatively higher than the interval centers and purple color represents the contrary. In most cases, the interval centers fall below the SKLM predictions. This behavior is expected as the depth-to-load conversions in the Utah Snow Load study were known to be relatively conservative, resulting in higher design ground snow load predictions than would be obtained by other methods \citep{Bean2018-report}. Thus, the set of depth-to-load conversion methods results in a set of predictions for which the SKLM approach represents an approximate upper bound. While conservative design snow load estimates are perhaps desirable on the context of building design, the interval-valued approach quantifies the degree of conservatism that results from using the 2018 Utah approach. For example, design ground snow load predictions in the major municipalities of the Wasatch Front (such as Logan and Salt Lake City Utah) tend to be 15\% to 25\% above the center of the range of predictions provided by the entire set of depth-to-load conversions. The ability to quantify the imprecision in the design ground snow load estimates at any point in the state could allow engineers to make dynamic adjustments to safety factors in a load resistance factor design (LRFD). In particular, LRFD safety factors could be increased in areas with low data precision, and reduced in areas with high data precision. The preservation of context resulting from these interval-valued predictions could be extended to other climate-based mapping problems where the data inputs are inherently imprecise.  

\begin{figure}
\centering
\includegraphics[width=0.6\textwidth]{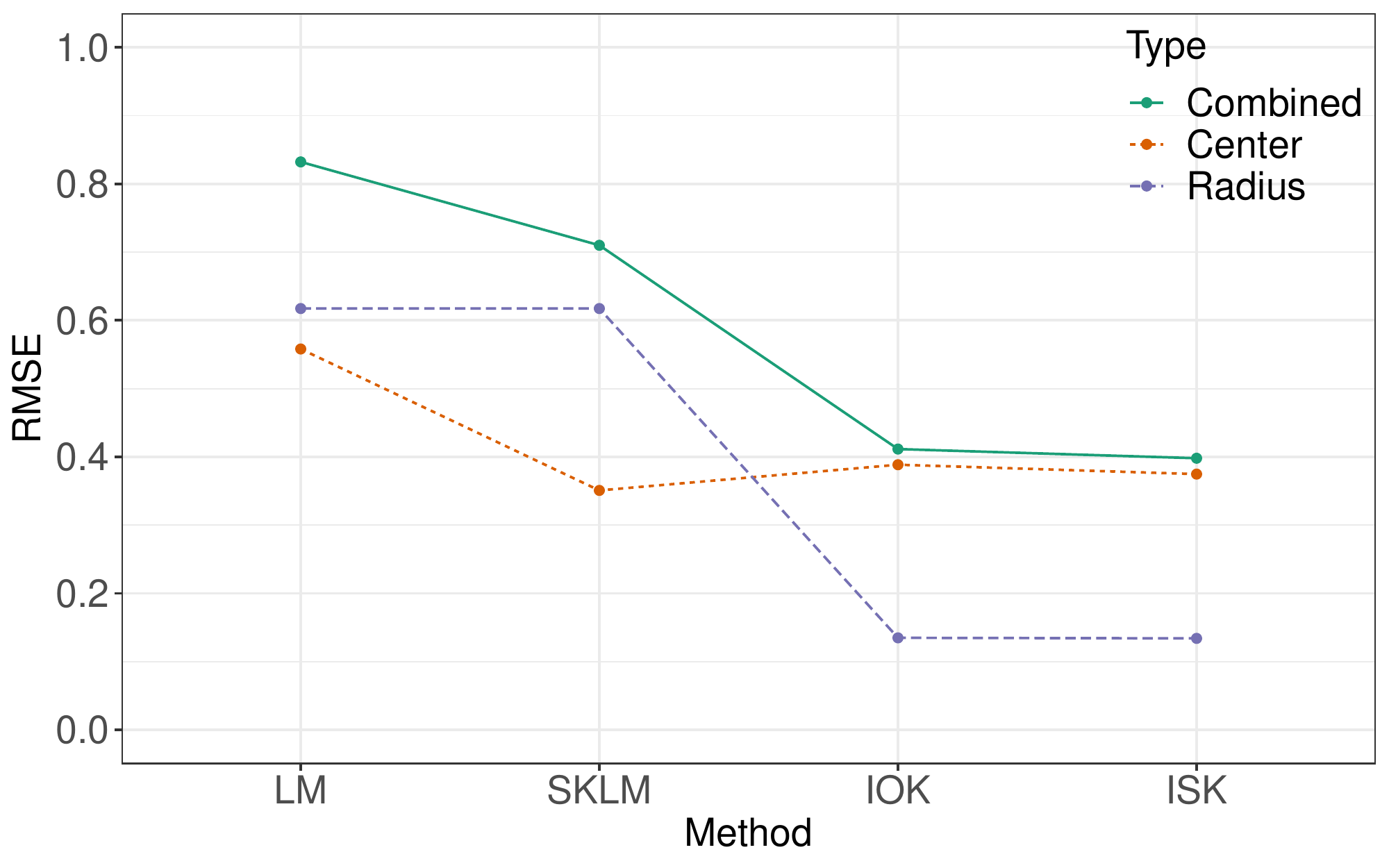}
\caption{Comparisons of the root mean squared error (RMSE) for the centers, radius, and intervals (combined) using a basic linear model approach (LM), the point-valued kriging approach (SKLM), and interval-valued versions of ordinary (IOK) and simple (ISK) kriging.}
\label{fig:cvSum}
\end{figure}

Figure \ref{fig:cvSum} illustrates the notion of accuracy in the interval-valued kriging framework via 10-fold cross validation. One of the difficulties in comparing interval-valued kriging to its point-valued counterparts is that traditional notions accuracy compare distances between points, rather than intervals. For example, the root mean squared error (RMSE) is defined according to the $\rho_2^2$ metric as
\[
\text{RMSE} = \sqrt{(1/n)\sum_i\left[(\hat{Y}^C - Y^C)^2 + (\hat{Y}^R - Y^R)^2\right]}.
\]
This is in contrast to traditional measures of accuracy for the centers and radius defined as 
\begin{eqnarray*}
\text{RMSE}(C) &=& \sqrt{(1/n)\sum_i(\hat{Y}^C - Y^C)^2} \\
\text{RMSE}(R) &=& \sqrt{(1/n)\sum_i(\hat{Y}^R - Y^R)^2}.
\end{eqnarray*}
This makes it difficult to say that the interval-valued approach is more or less accurate than its point-valued counterparts. The comparative problem cannot be solved by simply making separate, point-valued predictions of center and radius. Mathematically, the allowance of negative weights in the kriging framework creates the possibility of negative-valued predictions for the interval-radius. Practically, separate predictions of the center and radius artificially separate the elements of a single interval observation. For this reason, we report and compare each defined RMSE metric for a model which only accounts for elevation and assumes no spatial component (LM), a regression-kriging model intended for predicting interval-centers (SKLM) and interval-valued ordinary (IOK) and simple (ISK) regression models. These results show that the SKLM predictions for the centers are slightly better in terms of cross validated error than the interval-valued approaches. This is to be expected as the interval-valued models must simultaneously minimize the prediction variance for both center and radius, which leads to slightly higher when compared to separate estimations of either center or radius. However, we also observe that this simultaneous consideration of center and radius leads to a reduction in the overall interval RMSE when compared to approaches that ignore one or more spatially dependent elements of the interval.  

\section{Conclusion}\label{reflections}
There are situations in which the efficacy of geostatistical model outputs rely on proper characterizations of the imprecision in the model inputs. One notable example of this is the imprecision introduced in design snow load estimates when depth-to-load estimations are required. We have demonstrated a new approach for handling such imprecision through our interval-valued kriging models, which simplify and clarify the earlier development in \cite{Diamond88}. These models use a real-valued covariance between intervals that allows for estimates of prediction uncertainty (via the kriging variance) \textit{and} prediction imprecision (via the interval radius), while retaining the feel and form of their point-valued counterparts. This makes our interval-valued kriging models a natural extension for handling input imprecision to anyone familiar with the kriging paradigm. Our results demonstrate the feasibility of interval-valued snow load predictions in the state of Utah and illustrate the preservation of valuable context that occurs when data imprecision is allowed to persist through the mapping technique. Further, we demonstrate through cross validation the improvement in interval-valued predictions that occurs when the spatial correlations for both centers and radii are considered in the kriging model.     

Although the center is a natural measure of level for an interval, it is not necessarily so in general. More rigorously, one needs to consider weighting within the interval and compute the level by means of an integral, e.g., $\int_{[0,1]}\left[\lambda s^L+(1-\lambda)s^U\right]\nu(d\lambda)$, where $\nu$ is a normalized measure on [0, 1] characterizing the weighting. Future work on our interval-valued kriging models will include the development of such methods for computing the level, which is usually of great interest in practice. We also anticipate adjustments to the optimization algorithm that will allow for direct use of the variogram in the numerical implementation, rather than assuming the existence of $C(\bdsm{0})$. In addition, considerations of uncertainty in the theoretical variogram parameters, similar to \cite{Loquin2012}, would be worth further investigation.  Indeed, one of the crucial products of this paper are the many future considerations made possible by a new mathematical foundation and computationally feasible implementation of interval-valued kriging.
We conclude that our interval-valued kriging models provide a practical and important alternative for researchers looking to extend their kriging applications to accommodate interval-valued inputs. These accommodations of imprecise inputs are imperative to ensuring that the spatial models we create today can meet the data challenges of tomorrow. 


\section{Appendix}
\subsection{Proofs}
\subsubsection{Proof of Theorem \ref{thm:kriging_var}}
\begin{proof}
The prediction variance is defined as 
\begin{eqnarray}
  &&E\left[\rho^2_K\left([\hat{Z}(x^*)], [Z(x^*)]\right)\right]\nonumber\\
  &=&E\big[A_{11}\left(\hat{Z}^C(x^*)-Z^C(x^*)\right)^2+A_{22}\left(\hat{Z}^R(x^*)-Z^R(x^*)\right)^2+ \nonumber \\
&& 2A_{12}\left(\hat{Z}^C(x^*)-Z^C(x^*)\right)\left(\hat{Z}^R(x^*)-Z^R(x^*)\right)\big]
  \nonumber\\
  &:=&A_{11}\text{I}+A_{22}\text{II}+2A_{12}\text{III}.\label{eqn:var_decomp}
\end{eqnarray}
First of all, by the unbiasedness of $\hat{Z}^C(\cdot)$,
\begin{eqnarray}
  \text{I}&=&E\left(\hat{Z}^C(x^*)-Z^C(x^*)\right)^2\nonumber\\
  &=&\text{Var}\left(\hat{Z}^C(x^*)-Z^C(x^*)\right)\nonumber\\
  &=&\text{Var}\left(\sum\lambda_iZ^C(x_i)\right)+\text{Var}\left(Z^C\right)-2\text{Cov}\left(\sum\lambda_iZ^C(x_i), Z^C(x^*)\right)\nonumber\\
  &=&\sum_i\sum_j\lambda_i\lambda_j\text{Cov}\left(Z^C(x_i), Z^C(x_j)\right)-2\sum_i\lambda_i\text{Cov}\left(Z^C(x_i), Z^C\right)+\text{Var}\left(Z^C(x^*)\right)\nonumber\\
  &=&\sum_i\sum_j\lambda_i\lambda_jC^{C,C}(x_i-x_j)-2\sum_i\lambda_iC^{C,C}(x_i-x^*)+C^{C,C}(0).\label{eqn:var_I}
\end{eqnarray}
Second, in the similar fashion, 
\begin{eqnarray}
  \text{II}&=&E\left(\hat{Z}^R(x^*)-Z^R(x^*)\right)^2  \nonumber \\
  &=& \sum_i\sum_j\left|\lambda_i\lambda_j\right|C^{R,R}(x_i-x_j)-2\sum_i\left|\lambda_i\right|C^{R,R}(x_i-x^*)+C^{R,R}(0).\label{eqn:var_II}
\end{eqnarray}
Finally, 
\begin{eqnarray}
\text{III}&=&E\left(\hat{Z}^C(x^*)-Z^C(x^*)\right)\left(\hat{Z}^R(x^*)-Z^R(x^*)\right)\nonumber\\
&=&\text{Cov}\left(\hat{Z}^C(x^*)-Z^C(x^*), \hat{Z}^R(x^*)-Z^R(x^*)\right)\nonumber\\
&=&\text{Cov}\left(\sum\lambda_iZ^C(x_i), \sum|\lambda_j|Z^R(x_j)\right)-\text{Cov}\left(Z^C(x^*), \sum|\lambda_i|Z^R(x_i)\right)\nonumber\\
&&-\text{Cov}\left(\sum\lambda_iZ^C(x_i), Z^R(x^*)\right)+\text{Cov}\left(Z^C(x^*), Z^R(x^*)\right)\nonumber\\
&=&\sum_i\sum_j\lambda_i|\lambda_j|C^{C,R}(x_i-x_j)-\sum_i|\lambda_i|C^{R,C}(x_i-x^*) \nonumber \\
&&-\sum_i\lambda_iC^{C,R}(x_i-x^*)+C^{C,R}(0).\label{eqn:var_III} 
\end{eqnarray}
Plugging (\ref{eqn:var_I})-(\ref{eqn:var_III}) into (\ref{eqn:var_decomp}) completes the proof. 
\end{proof}

\subsubsection{Proof of Theorem \ref{thm:kriging_var_vario}}
\begin{proof}
Recall (\ref{eqn:var_decomp}) in the proof of Theorem \ref{thm:kriging_var} that 
\begin{equation}\label{eqn:0}
  E\left[\rho^2_K\left([\hat{Z}(x^*)], [Z(x^*)]\right)\right]
  =A_{11}\text{I}+A_{22}\text{II}+2A_{12}\text{III}.
\end{equation}
From (\ref{eqn:vario-cov-c})-(\ref{eqn:vario-cov-cr2}), we have
\begin{eqnarray}
  C^{C,C}(\bdsm{h})&=&C^{C,C}(\bdsm{0})-\gamma^C(\bdsm{h}),\label{eqn:cov-vario-c}\\
  C^{R,R}(\bdsm{h})&=&C^{R,R}(\bdsm{0})-\gamma^R(\bdsm{h}),\label{eqn:cov-vario-r}\\
  C^{C,R}(\bdsm{h})+C^{C,R}(-\bdsm{h})&=&2\left[C^{C,R}(\bdsm{0})-\gamma^{C,R}(\bdsm{h})\right].\label{eqn:cov-vario-cr}
\end{eqnarray} 
Plugging (\ref{eqn:cov-vario-c}) in (\ref{eqn:var_I}) and by the unbiasedness constraints, we obtain
\begin{eqnarray}
  I=-\sum_i\sum_j\lambda_i\lambda_j\gamma^C(x_i-x_j)+2\sum_i\lambda_i\gamma^C(x_i-x^*).\label{eqn:I}
\end{eqnarray}
Similarly, by plugging (\ref{eqn:cov-vario-r}) in (\ref{eqn:var_II}), we get
\begin{eqnarray}
  II=-\sum_i\sum_j\lambda_i\lambda_j\gamma^R(x_i-x_j)+2\sum_i\lambda_i\gamma^C(x_i-x^*).\label{eqn:II}
\end{eqnarray}
Finally, the last term can be rewritten as
\begin{eqnarray}
  III&=&\left(\sum_i\lambda_i^2\right)C^{C,R}(\bdsm{0})+\sum_{i<j}\lambda_i\lambda_j\left[C^{C,R}(x_i-x_j)+C^{C,R}\left(-(x_i-x_j)\right)\right]\nonumber\\
  &&-\sum_i\lambda_i\left[C^{C,R}(x_i-x^*)+C^{C,R}\left(-(x_i-x^*)\right)\right]+C^{C,R}(\bdsm{0}).\label{eqn:var_III_alter}
\end{eqnarray}
Plugging (\ref{eqn:cov-vario-cr}) in (\ref{eqn:var_III_alter}), we get
\begin{eqnarray}
  III&=&\left(\sum_i\lambda_i^2\right)C^{C,R}(\bdsm{0})+2\sum_{i<j}\lambda_i\lambda_j\left[C^{C,R}(\bdsm{0})-\gamma^{C,R}(x_i-x_j)\right]\nonumber\\
  &&-2\sum_i\lambda_i\left[C^{C,R}(\bdsm{0})-\gamma^{C,R}(x_i-x^*)\right]+C^{C,R}(\bdsm{0})\nonumber\\
  &=&C^{C,R}(\bdsm{0})-2\sum_{i<j}\lambda_i\lambda_j\gamma^{C,R}(x_i-x_j)-2C^{C,R}(\bdsm{0})+2\sum_i\lambda_i\gamma^{C,R}(x_i-x^*)+C^{C,R}(\bdsm{0})\nonumber\\
  &=&-\sum_{i\ne j}\lambda_i\lambda_j\gamma^{C,R}(x_i-x_j)+2\sum_i\lambda_i\gamma^{C,R}(x_i-x^*).\label{eqn:III}
\end{eqnarray}
Plugging (\ref{eqn:I}), (\ref{eqn:II}), and (\ref{eqn:III}) in (\ref{eqn:0}) completes the proof. 
\end{proof}

\subsubsection{Proof of Theorem \ref{thm:Gradient-Hessian}}
\begin{proof}
Define
\begin{align}
  V^{C,C}(\bdsm{\lambda})
  &=\sum_i\sum_j\lambda_i\lambda_jC^{C,C}(\bdsm{x}_i-\bdsm{x}_j)-2\sum_i\lambda_iC^{C,C}(\bdsm{x}_i-\bdsm{x}^*),\label{eq:VC}\\
  V^{R,R}(\bdsm{\lambda})
  &=\sum_i\sum_j\left|\lambda_i\lambda_j\right|C^{R,R}(\bdsm{x}_i-\bdsm{x}_j)-2\sum_i\left|\lambda_i\right|C^{R,R}(\bdsm{x}_i-\bdsm{x}^*),\label{eq:VR}\\
  V^{C,R}(\bdsm{\lambda})
  &=\sum_i\sum_j\lambda_i|\lambda_j|C^{C,R}(\bdsm{x}_i-\bdsm{x}_j)-\sum_i|\lambda_i|C^{R,C}(\bdsm{x}_i-\bdsm{x}^*) \nonumber \\
  &-\sum_i\lambda_iC^{C,R}(\bdsm{x}_i-\bdsm{x}^*).\label{eq:VCR}
\end{align}
The prediction variance is then rewritten as 
\[
  E\left[\rho^2_K\left([\hat{Z}(\bdsm{x}^*)], [Z(\bdsm{x}^*)]\right)\right]
  =A_{11}V^{C,C}(\bdsm{\lambda}) + A_{22}V^{R,R}(\bdsm{\lambda}) + 2A_{12}V^{C,R}(\bdsm{\lambda}).
 \]
It follows that 
\begin{eqnarray}
 Q(\lambda) &=& A_{11}V^{C,C}(\bdsm{\lambda}) + A_{22}V^{R,R}(\bdsm{\lambda}) + 2A_{12}V^{C,R}(\bdsm{\lambda}) + P(\bdsm{\lambda}, c),
\label{eqn:Q1} 
\end{eqnarray}
and
\begin{eqnarray}
\bdsm{G} =\nabla Q(\lambda) &=& A_{11}(\partial/\partial\lambda_k)V^{C,C}(\bdsm{\lambda}) + A_{22}(\partial/\partial\lambda_k)V^{R,R}(\bdsm{\lambda}) + 2A_{12}(\partial/\partial\lambda_k)V^{C,R}(\bdsm{\lambda})  \nonumber \\
&& + (\partial/\partial\lambda_k)P(\bdsm{\lambda}, c) \quad  (k = 1, \ldots, n) \label{eqn:Qgrad} \\
 \bdsm{H}= \nabla^2 Q(\lambda) &=& A_{11}(\partial^2/\partial\lambda_k\lambda_l)V^{C,C}(\bdsm{\lambda}) + A_{22}(\partial^2/\partial\lambda_k\lambda_l)V^{R,R}(\bdsm{\lambda}) + 2A_{12}(\partial^2/\partial\lambda_k\lambda_l)V^{C,R}(\bdsm{\lambda})  \nonumber \\ 
& & + (\partial^2/\partial\lambda_k\lambda_l)P(\bdsm{\lambda}, c)  \quad(k, l = 1, \ldots, n) \label{eqn:Qhess} .
 \end{eqnarray} 

For SK, we use the quadratic approximation in (\ref{eqn:approx}) to handle the non-differentiability of $|\lambda|$, which amounts to
\begin{eqnarray*}
  (d/d\lambda)|\lambda| = (\lambda/\lambda_0),\ \ \ \ 
  (d^2/d\lambda^2)|\lambda| = (1/\lambda_0),\ \ \ \ 
  \text{for}\ \lambda\approx\lambda_0\neq 0. 
\end{eqnarray*}

Then, the components of $\nabla Q$ are calculated as
\begin{eqnarray*}
(\partial/\partial\lambda_k)V^{C,C} &=& \sum_i\lambda_i\left[C^{C,C}(\bdsm{x}_i - \bdsm{x}_k) + C^{C,C}(\bdsm{x}_k - \bdsm{x}_i)\right] - 2C^{C,C}(\bdsm{x}_k - \bdsm{x}^*), \\
(\partial/\partial\lambda_k)V^{R,R} &=& (\lambda_k/|\lambda_{k0}|)\left[\sum_{i}\left|\lambda_i\right|\left[C^{R,R}(\bdsm{x}_i - \bdsm{x}_k) + C^{R,R}(\bdsm{x}_k - \bdsm{x}_i)\right] - 2C^{R,R}(\bdsm{x}_k - \bdsm{x}^*)\right] \\
&=& (\lambda_k/|\lambda_{k0}|)f^{R,R}\left(\bdsm{\lambda}, \lambda_k\right), \\
(\partial/\partial\lambda_k)V^{C,R} &=& \sum_i \left[|\lambda_i|C^{C,R}(\bdsm{x}_k - \bdsm{x}_i) + (\lambda_k/|\lambda_{k0}|)\lambda_iC^{C,R}(\bdsm{x}_i - \bdsm{x}_k)\right] \\
&& - (\lambda_k/|\lambda_{k0}|)C^{R,C}(\bdsm{x}_k - \bdsm{x}^*) - C^{C,R}(\bdsm{x}_k - \bdsm{x}^*), \\
(\partial/\partial\lambda_k)P(\bdsm{\lambda}, c) &=& -(2c\lambda_k/|\lambda_{k0}|)\left(1-\sum_i|\lambda_i|\right). 
\end{eqnarray*}
Similarly, the components of $\nabla^2 Q$ are
\begin{eqnarray*}
&&(\partial^2/\partial\lambda_k\lambda_l)V^{C,C} = C^{C,C}(\bdsm{x}_l - \bdsm{x}_k) + C^{C,C}(\bdsm{x}_k - \bdsm{x}_l),  \\ 
&&(\partial^2/\partial\lambda_k\lambda_l)V^{R,R} = 
\begin{cases}
(1/|\lambda_{k0}|)f^{R,R}\left(\bdsm{\lambda}, \lambda_k\right) + 2C^{R,R}(\bdsm{0})\left((\lambda_k/|\lambda_{k0}|)\right)^2 & k = l \\
(\lambda_k\lambda_l/|\lambda_{k0}\lambda_{l0}|)\left[C^{R,R}(\bdsm{x}_l - \bdsm{x}_k) + C^{R,R}(\bdsm{x}_k - \bdsm{x}_l)\right] & k \ne l
\end{cases}, \\
\\
&&(\partial^2/\partial\lambda_k\lambda_l)V^{C,R} = 
\begin{cases}
(2\lambda_k/|\lambda_{k0}|)C^{C,R}(\bdsm{0}) + (1/|\lambda_{k0}|)\left[\sum_i\lambda_iC^{C,R}(\bdsm{x}_i - \bdsm{x}_k) - C^{R,C}(\bdsm{x}_k - \bdsm{x}^*)\right] & k = l \\
(\lambda_l/|\lambda_{l0}|)C^{C,R}(\bdsm{x}_k - \bdsm{x}_l) + (\lambda_k/|\lambda_{k0}|)C^{C,R}(\bdsm{x}_l - \bdsm{x}_k) & k \ne l
\end{cases}, \\
\\
&&(\partial^2/\partial\lambda_k\lambda_l)P(\bdsm{\lambda}, c) =
 \begin{cases} 
2c\left((\lambda_k/|\lambda_{k0}|)\right)^2 - (2c/|\lambda_{k0}|)\left(1-\sum_i|\lambda_i|\right)  & k = l \\
(2c\lambda_k\lambda_l/|\lambda_{k0}||\lambda_{l0}|)& k \ne l
\end{cases}.
\end{eqnarray*}
Plugging these derivations into (\ref{eqn:Q1} - \ref{eqn:Qhess}) completes the SK case. 

For OK, under the constraint that $\lambda_i \ge 0$, we have $|\lambda_i|=\lambda_i$ and $Q(\bdsm{\lambda})$ is a quadratic function, so no approximation is needed. The components of the gradient and Hessian are calculated straightforwardly as
\begin{eqnarray*}
(\partial/\partial\lambda_k)V^{C,C} &=& \sum_i\lambda_i\left[C^{C,C}(\bdsm{x}_i - \bdsm{x}_k) + C^{C,C}(\bdsm{x}_k - \bdsm{x}_i)\right] - 2C^{C,C}(\bdsm{x}_k - \bdsm{x}^*), \\
(\partial/\partial\lambda_k)V^{R,R} &=& \sum_{i}\lambda_i\left[C^{R,R}(\bdsm{x}_i - \bdsm{x}_k) + C^{R,R}(\bdsm{x}_k - \bdsm{x}_i)\right] - 2C^{R,R}(\bdsm{x}_k - \bdsm{x}^*), \\
(\partial/\partial\lambda_k)V^{C,R} 
&=& \sum_i \lambda_i\left[C^{C,R}(\bdsm{x}_k - \bdsm{x}_i) + C^{C,R}(\bdsm{x}_i - \bdsm{x}_k)\right] \\
&& - C^{R,C}(\bdsm{x}_k - \bdsm{x}^*) - C^{C,R}(\bdsm{x}_k - \bdsm{x}^*), \\
(\partial/\partial\lambda_k)P(\bdsm{\lambda}, c) &=& -(2/c)\left(1-\sum_i\lambda_i\right) - (c/\lambda_k),  \\
(\partial^2/\partial\lambda_k\partial\lambda_l)V^{C,C} &=& C^{C,C}(\bdsm{x}_l - \bdsm{x}_k) + C^{C,C}(\bdsm{x}_k - \bdsm{x}_l),  \\ 
(\partial^2/\partial\lambda_k\lambda_l)V^{R,R} &=& 
C^{R,R}(\bdsm{x}_l - \bdsm{x}_k) + C^{R,R}(\bdsm{x}_k - \bdsm{x}_l), \\
(\partial^2/\partial\lambda_k\lambda_l)V^{C,R} &=& 
C^{C,R}(\bdsm{x}_k - \bdsm{x}_l) + C^{C,R}(\bdsm{x}_l - \bdsm{x}_k), \\
(\partial^2/\partial\lambda_k\lambda_l)P(\bdsm{\lambda}, c) &= &
(2/c) + (c/\lambda_k^2)I_{(k=l)}.
\end{eqnarray*}
Plugging these results into (\ref{eqn:Q1} - \ref{eqn:Qhess}) completes the OK case. 
\end{proof}

\subsection{Methods for estimating snow depth from snow load}
Let $q_s$ represent the weight of snow on the ground, also called the ground snow load. $q_s$ is often estimated as a function of snow depth $h$, measured in centimeters (cm). The following techniques describe the way in which three mountainous states estimate the weight of snow from its depth. 

\subsubsection{Colorado's Method}
Colorado converts snow depth to snow load using two non-linear curves, one created by \cite{Tobiasson1997} and another developed by \cite{Debock2017}. These curves are defined as 
\[
q_s = g_1(h, A(\bdsm{x})) = p(A(\bdsm{x}))*f^{(1)}(h) + (1-p(A(\bdsm{x})))*f^{(2)}(h)
\]
with
\[
f^{(1)} = (0.0479)(0.279)\left(\frac{h}{2.54}\right)^{1.36} \qquad
f^{(2)} = (0.0479)(0.584)\left(\frac{h}{2.54}\right)^{1.15}. 
\]
The load parameter $p \in \left[0, 1\right]$ reaches its lower and upper limits for elevations ($A$) of around 1800 m and 2600 m respectively \citep{SEAC2016}. 

\subsubsection{Idaho's Method}
Idaho uses the Rocky Mountain Conversion Density (RMCD) \citep{Sack1986} redefined for metric units as
\[
q_s(h) = g_2(h) = \begin{cases}
0.017h & h< 55.88\text{cm} \\
0.0445h-1.5274 & h\ge 55.88\text{cm}
\end{cases}
\]
where $h$ represents snow depth (cm) \citep{Sack2015-2}. Note that this method, as well as Colorado's approach, are only designed to predict the maximum annual snow load using the maximum annual snow depth. 

\subsubsection{Utah's Method}
Utah uses a snow density estimation model created by \cite{Sturm2010} and referred to hereafter as ``Sturm's equation.'' Unlike the previous methods which focus only on modeling annual maximum snow depths, Sturm's equation is designed for depth-to-load conversions on any day of the snow season. Assuming the maximum density of water, this method models snow load with the equation
\[
q_s = g_3(h, d) =  0.0981h\left[\left(\rho_{max}- \rho_0\right)\left[1 - \exp\left(-k_1h - k_2d\right)\right] + \rho_0\right]
\]
where $d$ represents day of the snow season starting on October 1st (-92) and ending June 30th (181) with no zero value. Additionally, $\rho_o, \rho_{max}, k_1, \text{ and } k_2$ are parameters specific to a particular climate class defined in Table 4 of \cite{Sturm2010} and provided for convenience in Table \ref{tbl:Sturm1} of this paper. 

\begin{table}
\centering
\caption{Climate specific parameters for Sturm's equation.}
\label{tbl:Sturm1}
\begin{tabular}{lcccc}
\textbf{Class} & $\rho_{max}$ & $\rho_{0}$ & $k_1$ & $k_2$ \\
Alpine & 0.5975 & 0.2237 & 0.0012 & 0.0038 \\ 
Maritime & 0.5979 & 0.2578 & 0.0010 & 0.0038 \\
Prairie & 0.5940 & 0.2332 & 0.016 & 0.0031 \\
Tundra & 0.3630 & 0.2425 & 0.0029 & 0.0049 \\
Taiga & 0.2170 & 0.2170 & 0.0000 & 0.0000 \\
\end{tabular}
\end{table}

\cite{Sturm1995} classifies nearly all of Utah as a ``Prairie'' climate type. However, the coarse resolution of their classification map (50 km by 50 km) makes it reasonable to believe that high elevation locations in Utah would likely be considered ``alpine'' if the grid was finer. Thus, the Utah study \citep{Bean2018-report} performed depth-to-load conversions for ``prairie'' and ``alpine'' terrains using the equation 
\[
q_s =  \begin{cases}
 0.0981h\left[.3608*\left(1 - \exp\left(-.0016h - .0031d\right)\right) + .2332\right] & A < 2113.6 \,\text{m} \\
 0.0981h\left[.3738*\left(1 - \exp\left(-.0012h - .0038d\right)\right) + .2237\right] & A >= 2113.6 \,\text{m}.
\end{cases}
\]

\bibliographystyle{elsarticle-harv}
\bibliography{masterReference}

\end{document}